\documentclass[aps,prd,preprint,groupedaddress]{revtex4}

\usepackage{graphicx}
\usepackage{dcolumn}
\usepackage{bm}
\usepackage{eufrak}

\def\({\left(}
\def\){\right)}
\def\[{\left[}
\def\]{\right]}
\def\be{\begin{equation}}
\def\ee{\end{equation}}
\def\bea{\begin{eqnarray}}
\def\eea{\end{eqnarray}}
\def\ba{\begin{array}}
\def\ea{\end{array}}
\def\pa{\partial}
\def\vr{\vec{r}}
\def\la{\lambda}
\def\vu{\vec{u}}
\def\na{\nabla}

\def\de{\delta}
\def\al{\alpha}
\def\a{\alpha}

\def\Y{\mathbf{Y}}
\def\q{\mathbf{q}}
\def\p{\mathbf{p}}
\def\b{\mathbf{b}}
\def\I{\mathbf{I}}
\def\h{\mathbf{h}}
\def\H{\mathbf{H}}
\def\0{\mathbf{0}}

\begin{document}


\title{
Gravitational wave detection by a spherical antenna: the angular sensitivity of resonators in the TIGA configuration and its variation with sidereal time and galactic longitude}

\author{Maria Alice Gasparini}
\email{alice.gasparini@physics.unige.ch}
\affiliation{D\'epartement de Physique Th\'eorique, Universit\'e de Gen\`eve,
24 quai Ernest-Ansermet, CH-1211 Gen\`eve 4\\}

\date{\today}

\begin{abstract}
Experimental projects using spherical antennas to detect gravitational waves are nowdays a concrete reality.
The main purpose of this paper is to give a possible way of interpreting output data from such a system. Responses of the five fundamental quadrupole modes and of the six resonators in TIGA collocations are shown as a function of the incoming direction of the incident wave. Then, for a source lying in the galactic plane, sidereal time and galactic longitude dependence is given. 
Thus, once a candidate source of gravitational waves is considered, we can exactly predict the resonators' response as a function of time.  
\end{abstract}


\maketitle

\section{Introduction}\label{intro}

Theoretical interest in spherical gravitational wave (GW) antennas dates back to the 1970s. Spherical antennas have a greater cross-section than a bar detectors of similar dimensions. More importantly, they have both omnidirectional and omnipolarization sensitivity, and also the potential to detect the direction of wave provenance.

Interest in experimental research into resonant spheres has increased over the past 15 years, and today spherical antennas are recognized to be the new generation of gravitational resonant detectors, to complement existing cylindrical antennas.
Two experiments are under way: MiniGRAIL in Leiden (Holland) \cite{minigrail} and The Graviton Project in Sau Paulo (Brasil) \cite{saupaulo}.

One of the main problems today in detecting gravitational waves with resonant spheres is to find the ``best'' location for amplifiers on the detector surface in order to optimize sensitivity. To date, no theoretical basis has been developed to provide a definitive solution to this problem.
Nevertheless many suggestions have recently been proposed, including the uncoupled transducer configuration by Zhou and Michelson in 1994 \cite{Z&M}, and the PHC configuration by Lobo and Serrano in 1996 \cite{L&S}; both are five-resonator configurations, the first including radial and tangential motion of the resonators, the second only radial motion. In addition, the TIGA configuration, proposed by Merkowitz and Johnson \cite{M&J} (1995), consists of six transducers moving radially, and it enjoys a peculiar symmetry which greatly simplifies equations. 

The general theoretical aspects of an ideal simple sphere interacting only with gravitational radiation have already been treated in the literature (see for example \cite{W&P} or \cite{Ash}). In the first part of this paper we will work out briefly some of them and obtain the energies stored in the five fundamental sphere quadrupole modes, as well as their dependence on the incident wave direction and polarization.\\ Taking into account of earth's rotation and the detector's location, we will translate the direction dependence into galactic longitude and sidereal time dependence for a source lying in the galactic plane.

We will review the general problem of the spherical detector with radial moving resonators on its surface, again in the case of high signal-to-noise ratio (SNR). We then focus our attention on the TIGA configuration, finding the oscillation amplitudes for the six transducers as a function of the incident GW direction.\\
With the same method used in the case of the simple sphere without amplifiers, we will find the resonators' amplitude dependence on sidereal time and on galactic longitude for a source lying in the galactic plane which emits randomly polarized GW radiation.

The paper is organized as follows. In section 2 we briefly sketch some general aspects of the physics of GW resonant detectors without any assumption on the shape of the antenna. In section 3 we focus our attention on the spherical detector, and we present the solutions to the equations of motion followed by the five modes angular and sidereal time sensitivity. In section 4 results of the previous sections are generalized to the case of the spherical detector coupled to a set of radial transducers.
Section 5 contains the summary and conclusions.

\section{Generic Resonant Detectors}
\subsection{The Equations of Motion}

We will work with a solid whose density is $\rho(\vr)$, where $\vr=(r_1 , \, r_2 , \, r_3)=(x,\, y,\, z)$ is the position of an infinitesimal mass element relative to the center of the solid and we call the Lam\'e coefficients of the material $\la$ and $\mu.$

We suppose that a gravitational wave hits our detector at the time $t=0$.
If we call $\vec{u}(\vr,t)$ the small displacement from the equilibrium position at the point $\vec{r}$ and time $t$, the equation of motion turns out to be \cite{LL}
\be\label{eom}
\rho\frac{\pa^2\vu(\vr,t)}{\pa t^2}-\mu\nabla^2\vu(\vr,t)-(\la+\mu)\vec{\na}(\vec{\na}\cdot\vu(\vr,t))=\vec{F}(\vr,t),
\ee
where $F^{j}(\vr,t)=\rho R^{j}_{0k0}x^{k}(\vr, t)$ is the $j^{th}$ component of the tidal force density attributable to the GW \cite{gravitation}.

As is well known for a resonant detector the spatial dependence of the Riemann tensor can be neglected. Moreover, as all our summation indices are spatial and we reasonably assume the background metric to be flat on the earth's surface ( $g_{\mu\nu}=\eta_{\mu\nu}+h_{\mu\nu}$ ), we can write them all as lowered indices. Thus, in the first order of $h$, the Riemann tensor can be written $R_{i0j0}=\frac{1}{2}\ddot{h}_{ij}(t),$ which is a rank 3 traceless symmetric tensor. It represents then a particle of spin 2, the graviton.\\
Such a tensor can be decomposed into a base of 5 matrices.
In order to do that, we follow the procedure used by Lobo \cite{Lobo95}, working with a basis of real matrices instead of imaginary matrices, as is usually done:

$$
M^{1c}=\sqrt{\frac{15}{16\pi}}\pmatrix{0&0&1\cr 0&0&0\cr 1&0&0};\;
M^{2c}=\sqrt{\frac{15}{16\pi}}\pmatrix{1&0&0\cr 0&-1&0\cr 0&0&0};
$$
$$
M^{1s}=\sqrt{\frac{15}{16\pi}}\pmatrix{0&0&0\cr 0&0&1\cr 0&1&0};\;
M^{2s}=\sqrt{\frac{15}{16\pi}}\pmatrix{0&1&0\cr 1&0&0\cr 0&0&0};
$$
\be\label{matrix}
M^{0}=\sqrt{\frac{5}{16\pi}}\pmatrix{-1&0&0\cr 0&-1&0\cr 0&0&2}.
\ee
Noting $I$ the identity matrix, the above matrices have the following properties:
\bea
 M^{\alpha}_{ij} M^{\beta}_{ij}=\frac{15}{8\pi}\delta_{\alpha\beta},\quad I_{ij} M^{\alpha}_{ij}=0,\nonumber\\
\frac{5}{2}I_{ij}I_{kl}+\sum_{\alpha} M^{\alpha}_{ij}M^{\alpha}_{kl}=\frac{15}{16\pi}(\de_{ik}\de_{jl}+\de_{il}\de_{jk}),
\eea
with $\alpha,\beta=0,1c,1s,2c,2s.$\\
Then we can rewrite the $j$ component of the force density:

\bea\label{fg}
&&F_j(\vr,t) =  \rho R_{j0k0}(t)r_k\nonumber\\ &&=\frac{\rho}{2}r_k\frac{16\pi}{15}R_{l0m0}\frac{15}{16\pi}(\de_{mj}\de_{lk}+\de_{lj}\de_{mk})\nonumber\\
&&=\frac{8\pi}{15} \sum_s \rho M^{\alpha}_{jk} r_k M_{lm}^{\alpha}R_{l0m0}=\sum_{\alpha} f^{\alpha}_j \ddot{h}^{\alpha}(t).
\eea

Repeated indices are summed. Eq.(\ref{fg}) introduces the functions $\ddot{h}^{\alpha}(t)=\frac{8\pi}{15}M_{lm}^{\alpha}R_{l0m0}$ and $f_j^{\alpha}(\vr)=\rho M^{\alpha}_{jk} r_k$.\\ 
Direct computation gives for the $\vec{f}^{\al}$:
\bea
 \vec{f}^{1c}(\vr)=\rho\sqrt{\frac{15}{16\pi}}\pmatrix{r_3\cr0\cr r_1} &\vec{f}^{1s}(\vr)=\rho\sqrt{\frac{15}{16\pi}}\pmatrix{0\cr r_3\cr r_2}\nonumber\\ 
\vec{f}^{2c}(\vr)=\rho\sqrt{\frac{15}{16\pi}}\pmatrix{r_1\cr-r_2\cr 0} &\vec{f}^{2s}(\vr)=\rho\sqrt{\frac{15}{16\pi}}\pmatrix{r_2\cr r_1\cr 0}&.\label{f}\nonumber\\ 
\vec{f}^{0}(\vr)=\rho\sqrt{\frac{5}{16\pi}}\pmatrix{-r_1\cr-r_2\cr 2r_3}
\eea

We now want to write down the explicit expressions for the functions $\ddot{h}^{\al}(t)$ given a GW propagating along a generic direction with respect to the detector coordinates system. \\
In the frame where the GW propagates along the $\hat{z}$ axis we can pass into the TT gauge in order to rewrite the Riemann tensor as
\be
R_{i0j0}=\frac{1}{2}\pmatrix{\ddot{h}_+&\ddot{h}_\times&0\cr\ddot{h}_\times&-\ddot{h}_+&0\cr0&0&0}_{ij} \, . 
\ee 
We call $\theta$ the angle from the $\hat{z}$ axis and $\phi$ the angle between $\hat{x}$ and the projection into the $\hat{x},\hat{y}$ plane. $(\hat{x},\hat{y},\hat{z})$ is the detector frame basis. For a generic arrival direction given by the angle $(\theta,\phi)$, we have to perform a rotation in order to obtain the Riemann in the antenna frame:
\be
R_{i0j0}=\frac{1}{2}\left(\mathcal{M}\pmatrix{\ddot{h}_+&\ddot{h}_\times&0\cr\ddot{h}_\times&-\ddot{h}_+&0\cr0&0&0}\mathcal{M}^T\right)_{ij}
\ee
with
\be
\mathcal{M}=\pmatrix{\cos(\theta)\cos(\phi)&-\sin(\phi)&\sin(\theta)\cos(\phi) \cr            \cos(\theta)\sin(\phi)&\cos(\phi)&\sin(\theta)\sin(\phi)\cr-\sin(\theta)&0&\cos(\theta)}\ .
\ee

We now are able to compute the components $\ddot{h}^{\al}(t)$ which are
\begin{widetext}
\bea\label{g}
\ddot{h}^0(t)&=&\sqrt{\frac{\pi}{5}}\sin^2(\theta)\ddot{h}_+(t)\nonumber\\
\ddot{h}^{1c}(t)&=&\sqrt{\frac{4\pi}{15}}\left(-\cos(\theta)\sin(\theta)\cos(\phi)\ddot{h}_+(t)+\sin(\theta)\sin(\phi)\ddot{h}_{\times}(t) \right)\nonumber\\
\ddot{h}^{1s}(t)&=&\sqrt{\frac{4\pi}{15}}\left( \cos(\theta)\sin(\theta)\sin(\phi)\ddot{h}_+(t)+\sin(\theta)\cos(\phi)\ddot{h}_{\times}(t)\right)\nonumber\\
\ddot{h}^{2c}(t)&=&
\sqrt{\frac{4\pi}{15}}\left(\left(1+\cos^2(\theta)\right)\left(\cos^2(\phi)-\frac{1}{2}\right)\ddot{h}_+(t)-2\cos(\theta)\sin(\phi)\cos(\phi)\ddot{h}_{\times}(t)\right)
\nonumber\\
\ddot{h}^{2s}(t)&=&\sqrt{\frac{4\pi}{15}}\left(\left(1+\cos^2(\theta)\right)\cos(\phi)\sin(\phi)\ddot{h}_+(t)+\cos(\theta)\left(\cos^2(\phi)-\sin^2(\phi)\right)\ddot{h}_{\times}(t) \right).\nonumber\\
\eea
\end{widetext}
Note that we have defined the $+$ and $\times$ polarization with respect to the detector frame. In the case where the source has an intrinsic polarization $e_+,e_{\times},$ we have to make the substitution
\bea\label{pol}
h_{+}&=&\cos(2\psi)e_{+}+\sin(2\psi)e_{\times}\nonumber\\
h_{\times}&=&-\sin(2\psi)e_{+}+\cos(2\psi)e_{\times}
\eea
where $\psi$ describes the polarization angle. This becomes useful if we want to take the average of the polarization.

\subsection{ The general solution}

The solution of Eq.(\ref{eom}) can be expressed formally by means of the Green function integral \cite{Lobo95}, getting:

\be \label{SolGenerale}
\vu(\vr,t)=\sum_{\alpha}\sum_{N}\omega_{N}^{-1}\mathfrak{f}_{N}^{\al}\mathfrak{g}_{N}^{\al}(t)\vec{\Phi}_{N}(\vr)
\ee

where

\bea\label{fgfrak}
\mathfrak{f}^{\al}_{N}&=&M^{-1}\int_{\rm{Solid}}\vec{\Phi}_{N}(\vr)\cdot\vec{f}^{\al}(\vr) d^3 r,\label{fn_alpha}\nonumber\\
\mathfrak{g}^{\al}_{N}(t)&=&\int_0^t \ddot{h}^{\al}(t')\sin\omega_{N}(t-t') dt'
\eea
and $\vec{\Phi}_{N}(\vr)$ are the normalized (i.e. they verifies $\int_{\rm{Solid}}\vec{\Phi}_{N}(\vr)\cdot\vec{\Phi}_{N'}(\vr)\rho(\vr) d^3r=M\de_{N,N'}$) eigenfunctions solutions of the corresponding homogeneous equation
\be\label{sta}
\rho\omega_{N}^2\vec{\Phi}_{N}+\mu\na^2\vec{\Phi}_{N}+(\la+\mu)\vec{\na}(\vec{\na}\cdot\vec{\Phi}_{N})=0 \label{equ_hom}
\ee
with the appropriate boundary conditions.
$N$ indicates a collective quantum number denoting the quantum state.
 
 Putting the above result in a different form, i.e. writing the solution as
\be\label{B}
\vu(\vr,t)=\sum_N B_N(t)\vec{\Phi}_N(\vr),
\ee
it is obvious, as expected from \cite {gravitation} that the quantity  $B_N=\sum_{\alpha}\omega_{N}^{-1}\mathfrak{f}^{\al}_{N}\mathfrak{g}^{\al}_{N}(t)$
satisfies equation
\bea\label{eomB}
&\ddot{B}_{N}(t)&+\omega_{N}^2 B_{N}(t)=F_{N}(t)=
\sum_{\alpha}\mathfrak{f}^{\al}_{N}\ddot{h}^{\al}(t)\nonumber \\
       &=&\sum_{\alpha}M^{-1}\int_{\rm{Solid}}\vec{\Phi}_{N}(\vr)\cdot\vec{f}^{\al}(\vr) \ddot{h}^{\al}(t)d^3 r\nonumber \\
       &=&\int_{\rm{Solid}}M^{-1}\rho(\vr) R_{j0k0}(t)r_k \Phi_{N,j}(\vr).
\eea

Each mode is then formally equivalent to a one-dimensional harmonic oscillator with frequency $\omega_{N}$, driven by a force per unit mass $F_{N}=\sum_{\alpha}\mathfrak{f}^{\al}_{N}\ddot{h}^{\al}(t).$
For such a system, the expression for the energy per mass unit adsorbed from the driving force is given by \cite{Landau}:
\be \label{NRJ}
E_s=\frac{1}{2}\arrowvert\int_{-\infty}^{+\infty}F_{N}(t)e^{-i\omega_{N}t}dt\arrowvert^2.
\ee
Before moving  on to the case of the spherical detector, it would be intersting to use this general method on a cylindrical detector. While we omit this exercise here, it can easily be verified that this general method reproduces the results for the eigenfunctions and the energy stored in oscillations in the well-known case of the one-dimensional homogeneous bar \cite{Thorne}.

\section{The spherical detector}

\subsection{The normal modes}

The normal eigenfunctions of Eq.(\ref{sta}) for the sphere can be found in the literature (see, for example \cite{Ash}). Note that they are usually  given in terms of imaginary spherical harmonics. Here we will put all results in terms of real spherical harmonics, but the shape of the solutions is the same.\\
Let us briefly recall the main points. There are two kinds of solutions with boundary conditions  $\sigma_{ij}n_j=0$ at $r=R$, with
$\sigma_{ij}$ the stress tensor, conventionally defined as $\sigma_{ij}=\lambda_{kk}\de{ij}+2\mu u_{ij}$, with $u_{ij}=\frac{1}{2}(u_{i,j}+u_{j,i}).$
These solutions are called the $\it{toroidal}$ ones and the $\it{spheroidal}$ ones: 
\bea
\vec{\Phi}_{nl\alpha}^T(\vr)&=&T_{nl}(r) i\vec{L} Y_{l\alpha}(\theta,\phi)\nonumber \\
\vec{\Phi}_{nl\alpha}^S(\vr)&=&A_{nl}(r)Y_{l\alpha}(\theta,\phi)\hat{n}-B_{nl}(r)i\hat{n}\times\vec{L} Y_{l\alpha}(\theta, \phi)
\eea
where each mode $n,l,\alpha$ corresponds to the generic $N$ in Eq.(\ref{sta}), $n$ is a positive integer which represents the energy level for a fixed angular momentum $l$ and $Y_{l\alpha}$ are the real spherical harmonics with kinetic momentum $l$ obtained by the imaginary ones $Y_{l,\pm m}$ in the following way:
\bea\label{sphehar}
Y_{l,0}&=&Y_{l,0} \nonumber \\
Y_{l,mc}&=&\frac{1}{\sqrt{2}}(Y_{l,-m}+(-1)^m Y_{l,+m}) \nonumber \\
Y_{l,ms}&=&\frac{i}{\sqrt{2}}(Y_{l,-m}+(-1)^{m+1}Y_{l,+m}) \nonumber \\
\eea
with $m=0,...,l.$ 
So defined, all the spherical harmonics (\ref{sphehar}) are real quantities. For $l=2$ this gives:
\bea
Y_{2,0}&=&\sqrt{\frac{5}{4\pi}}\left(\frac{3}{2}\cos^2\theta-\frac{1}{2} \right) \nonumber \\
Y_{2,1c}&=&\sqrt{\frac{15}{4\pi}}\frac{1}{2}\sin2\theta\cos\phi \nonumber \\
Y_{2,1s}&=&\sqrt{\frac{15}{4\pi}}\frac{1}{2}\sin2\theta\sin\phi \nonumber \\
Y_{2,2c}&=&\sqrt{\frac{15}{4\pi}}\frac{1}{2}\sin^2\theta\cos2\phi \nonumber \\
Y_{2,2s}&=&\sqrt{\frac{15}{4\pi}}\frac{1}{2}\sin^2\theta\sin2\phi.
\eea
Also, given a generic unity vector in spherical coordinates, $\hat{n}=(\sin\theta\cos\phi,\sin\theta\sin\phi,\cos\theta)$ we can see that
\be
Y_{2,\al}(\theta,\phi)=M_{ij}^{\al}n_in_j
\ee  
where $M_{ij}^{\al}$ are the matrices defined in Eq.(\ref{matrix}).

$A_{nl}(r),$ $B_{nl}(r)$ and $T_{nl}(r)$ are scalar functions of $r$:

\bea
T_{nl}(r)&=&C'(n,l)j_l(k_{nl}r)\; ,\nonumber\\
A_{nl}(r)&=&C(n,l)\left[\beta_3(k_{nl}R)j_l'(q_{nl}r)\right.\nonumber\\&&\left.-l(l+1)\frac{q_{nl}}{k_{nl}}\beta_1(q_{nl}R)\frac{j_l (k_{nl}r)}{k_{nl}r}   \]\; ,\nonumber\\
B_{nl}(r)&=&C(n,l)\left[\beta_3(k_{nl}R)\frac{j_l(q_{nl}r)}{q_{nl}r}\right.\nonumber\\&&\left.-\frac{q_{nl}}{k_{nl}}\beta_1(q_{nl}R)\frac{[k_{nl}r j_l (k_{nl}r)]'}{k_{nl}r}   \]
\eea
where the $j_l$ are spherical Bessel functions and
\bea
\beta_0(x)&=&\frac{j_l(x)}{x^2},\quad\beta_1(x)=\frac{d}{dx}\frac{j_l(x)}{x},\quad\beta_2(x)=\frac{d^2}{dx^2}j_l(x),\nonumber\\
\beta_3(x)&=&\frac{1}{2}\beta_2(x)+\[\frac{l(l+1)}{2}-1\]\beta_0(x).
\eea
The $k$ and the $q$ are related to the quantized $\omega$ of Eq.(\ref{sta}) by
\be
k_{nl}^2=\frac{\rho\omega_{nl}^2}{\mu}\quad q_{nl}^2=\frac{\rho\omega_{nl}^2}{\lambda+2\mu}.
\ee

\subsection{Interaction with a gravitational wave}

We start by computing $\mathfrak{f}^{\al}_{nl\alpha'}$ of Eq.(\ref{fgfrak}) for our spherical detector for both toroidal and spheroidal eigenfunctions:
\bea\label{an}
\mathfrak{f}^{\al S}_{nl\alpha'}
&=&M^{-1}\int_{Sphere}\vec{\Phi}^S_{nl\alpha'}(\vr)\cdot\vec{f}^{\al}(\vr)d^3r=a_n\de_{l,2}\de_{\alpha,\alpha'}\nonumber\\
\mathfrak{f}^{\al T}_{nl\alpha'}
&=&M^{-1}\int_{Sphere}\vec{\Phi}^T_{nl\alpha'}(\vr)\cdot\vec{f}^{\al}(\vr)d^3r=0,
\eea
with
\be\label{a_n}
a_n=-\frac{1}{M}\int_0^R\rho r^3\left(A_{n2}(r)+3B_{n2}(r)\right)dr.
\ee
It is clear from Eq.(\ref{an}) that toroidal modes do not enter into play, so we will work from now on only with spheroidal modes and we will drop the $S$ for simplicity.

Once the Riemann tensor and the $\ddot{h}^{\al}(t)$ are known for a gravitational wave coming from a generic direction $(\theta,\phi)$, it is easy to compute $\mathfrak{g}^{\al}_{nl\alpha'}(t)$ using Eq.(\ref{fgfrak}). The solution of Eq.(\ref{eom}) can be written:
\be
\vu(\vr,t)=\sum_{n=1}^{\infty}\vu_n(\vr,t)=\sum_{n=1}^{\infty}\frac{a_n}{\omega_{n2}}\sum_{\alpha}\vec{\Phi}_{n2\alpha}(\vr)\mathfrak{g}^{\al}_{n2}(t).
\ee
For spheroidal modes, $B_N(t)=B_{n,l,\alpha}(t),$ as defined in the previous paragraph. In this case, $$B_N(t)=B_{n,l,\alpha}(t)=\sum_{\beta}\omega_{nl}^{-1}\mathfrak{f}_{n,l,\alpha}^{(\beta)}\mathfrak{g}_{n,l,\alpha}^{(\beta)}=\omega_{n2}^{-1}a_n\mathfrak{g}_{n2\alpha}^{\al},$$
and its equation of motion (\ref{eomB}) is
\be\label{Beomsf}
\ddot{B}_{n,l,\alpha}+\omega_{n2}^2B_{n,l,\alpha}=\de_{l,2}a_n\ddot{h}_{\al}(t).
\ee

The energy stored in the mode $\alpha$ at frequency $\omega_{nl}$ is (from Eq.(\ref{NRJ})):
\be\label{NRJS}
E_s(n,l,\alpha)=\frac{1}{2}\arrowvert\int_{-\infty}^{+\infty}F_{nl\alpha}(t) e^{-i\omega_{nl}t}dt\arrowvert^2.
\ee
Using Eqs.(\ref{eomB}) and (\ref{an}), 
Eq.(\ref{NRJS}) becomes
\be
E_s(n,l,\alpha)=\frac{1}{2}a_n^2\de_{l,2} \arrowvert \int_{-\infty}^{+\infty}\ddot{h}^{(\alpha)}(t)e^{-i\omega_{n2}t}dt  \arrowvert^2.
\ee
The only contributions have, as expected, $l=2.$
For these, we fix the value of $n$ 
and replace $\ddot{h}^{\alpha}$ using Eq.(\ref{g}). 

For the $\alpha=0$ mode, we have
\bea
\int_{-\infty}^{+\infty}\ddot{h}^{(0)}(t)&&e^{-i\omega_{n2}t}dt =\int_{-\infty}^{+\infty}\sqrt{\frac{\pi}{5}}\sin^2\theta\ddot{h}_{+}(t)e^{-i\omega_{n2}t}dt\nonumber\\&=&\sqrt{\frac{\pi}{5}}\sin^2\theta\omega^2_{n2}\int_{-\infty}^{+\infty}h_{+}(t)e^{-i\omega_{n2}t}dt\nonumber\\&=&\sqrt{\frac{\pi}{5}}\sin^2\theta\omega^2_{n2}\tilde{h}_+(\omega_{n2}),
\eea
thus
\be
E_s(n,2,0)=\frac{\pi}{10}a_n^2\sin^4\theta\omega^4_{n2}\arrowvert\tilde{h}_+(\omega_{n2})\arrowvert^2.
\ee

Similarly, for the other modes, we get:
\begin{widetext}
\bea\label{Ex}
E_s(n,2,1c)&=&\frac{2\pi}{15}a_n^2\omega^4_{n2}[ \cos^2\theta\sin^2\theta\cos^2\phi\arrowvert\tilde{h}_+\arrowvert^2+\sin^2\theta\sin^2\phi\arrowvert\tilde{h}_{\times}\arrowvert^2
-\cos\theta\sin^2\theta\sin2\phi \,  \Re( \tilde{h}_+\tilde{h}^*_{\times})  ],
\nonumber\\
E_s(n,2,1s)&=&\frac{2\pi}{15}a_n^2\omega^4_{n2}[ \cos^2\theta\sin^2\theta\sin^2\phi\arrowvert\tilde{h}_+\arrowvert^2+\sin^2\theta\cos^2\phi\arrowvert\tilde{h}_{\times}\arrowvert^2-
\cos\theta\sin^2\theta\sin2\phi\,\Re( \tilde{h}_+\tilde{h}^*_{\times}) ],
\nonumber\\
E_s(n,2,2c)&=&\frac{2\pi}{15}a_n^2\omega^4_{n2}[
(1+\cos^2\theta)^2(\cos^2\phi-\frac{1}{2})^2\arrowvert\tilde{h}_+\arrowvert^2+4\cos^2\theta\sin^2\phi\cos^2\phi\arrowvert\tilde{h}_{\times}\arrowvert^2\nonumber\\&-&4\cos\theta\sin\phi\cos\phi(1+\cos^2\theta)(\cos^2\phi-\frac{1}{2})\,\Re( \tilde{h}_+\tilde{h}^*_{\times})  ],
\nonumber\\
E_s(n,2,2s)&=&\frac{2\pi}{15}a_n^2\omega^4_{n2}[
(1+\cos^2\theta)^2\cos^2\phi\sin^2\phi\arrowvert\tilde{h}_+\arrowvert^2+\cos^2\theta(\cos^2\phi-\sin^2\phi)^2\arrowvert\tilde{h}_{\times}\arrowvert^2\nonumber\\&+&2\cos\theta\sin\phi\cos\phi(1+\cos^2\theta)(\cos^2\phi-\sin^2\phi)\,\Re( \tilde{h}_+\tilde{h}^*_{\times})  ].\nonumber\\
\eea

We have assumed that ${h}_{+,\times}(t)$ and $\dot{h}_{+,\times}(t)$ decrease fast enough at $\pm\infty$ to allow the integration by parts, $\tilde{h}_+(f)$ and $\tilde{h}_{\times}(f)$ stand for the Fourier transform of ${h}_+(t)$ and ${h}_{\times}(t).$ In this paper we use the notation $\tilde{a}(\omega)=\int_{-\infty}^{+\infty}dt a(t)e^{-i\omega t}$ for the Fourier transform and $a(t)=\int_{-\infty}^{+\infty}d\omega \tilde{a}(\omega)e^{+i\omega t}$ for the inverse. In Eq.(\ref{Ex}) all Fourier transforms are evaluated at the frequency  $f_{n2}=\omega_{n2}/2\pi$.

If the source emits randomly polarized radiation, we can rewrite $h_{\times}$ and $h_{+}$ in terms of one possible intrinsic polarization of the source (Eq.(\ref{pol})) and average over $\psi$. We finally obtain

\bea\label{nrj}
&&\bar{E}_s(n,2,0)=\frac{\pi}{15}a_n^2\omega^4_{n2}(\arrowvert\tilde{e}_+\arrowvert^2+\arrowvert\tilde{e}_{\times}\arrowvert^2)\frac{3}{4}\sin^4\theta,\nonumber\\
\nonumber\\
&&\bar{E}_s(n,2,1c)=\frac{\pi}{15}a_n^2\omega^4_{n2}(\arrowvert\tilde{e}_+\arrowvert^2+\arrowvert\tilde{e}_{\times}\arrowvert^2)\sin^2\theta(\cos^2\theta\cos^2\phi+\sin^2\phi),
\nonumber\\
&&\bar{E}_s(n,2,1s)=\frac{\pi}{15}a_n^2\omega^4_{n2}(\arrowvert\tilde{e}_+\arrowvert^2+\arrowvert\tilde{e}_{\times}\arrowvert^2)\sin^2\theta(\cos^2\theta\sin^2\phi+\cos^2\phi),
\nonumber\\
&&\bar{E}_s(n,2,2c)=\frac{\pi}{15}a_n^2\omega^4_{n2}(\arrowvert\tilde{e}_+\arrowvert^2+\arrowvert\tilde{e}_{\times}\arrowvert^2)[(1+\cos^2\theta)^2(\cos^2\phi-\frac{1}{2})^2+4\cos^2\theta\sin^2\phi\cos^2\phi],
\nonumber\\
&&\bar{E}_s(n,2,2s)=\frac{\pi}{15}a_n^2\omega^4_{n2}(\arrowvert\tilde{e}_+\arrowvert^2+\arrowvert\tilde{e}_{\times}\arrowvert^2)[(1+\cos^2\theta)^2\cos^2\phi\sin^2\phi+\cos^2\theta(\cos^2\phi-\sin^2\phi)^2].
\,
\nonumber\\
\eea
\end{widetext}
As before, all Fourier transforms are evaluated at the frequency $\omega_{n2}.$

It is important to stress at this point that if we sum the energies of Eq.(\ref{nrj}) we obtain a constant value:
\be
\bar{E}_s(n,2,tot)=\frac{2\pi}{15}a_n^2\omega^4_{n2}(\arrowvert\tilde{e}_+(\omega_{2n})\arrowvert^2+\arrowvert\tilde{e}_{\times}(\omega_{2n})\arrowvert^2).
\ee
This means that the total sensitivity of our detector is always the same, independently of the GW direction. This is to be expected because of the symmetry of the detector, and it is one of the most important properties differentiating spherical detectors from cylindrical ones.
Together with its greater sensitivity (keeping the same dimensions), this feature makes the new generation of resonant detectors much more advantageous than bars.

Figure 1 shows, for each mode $n$, how the energy stored in each $\alpha$ varies as a function of $\theta $ and $\phi,$ the angles that give the direction of the incoming unpolarized GW. The rate $$\epsilon_{\alpha}=\frac{15\bar{E}_s(n,2,\alpha)}{\pi a_n^2\omega_{n2}^4(\arrowvert\tilde{e}_+\arrowvert^2+\arrowvert\tilde{e}_{\times}\arrowvert^2)}$$ is zero where the graph is blue, and maximal ($\frac{3}{4}$) where the graph is red.

\begin{figure}\label{fig1}
\begin{center}
\begin{minipage}{4.5cm}
\includegraphics[width=4.5cm,height=4.5cm]{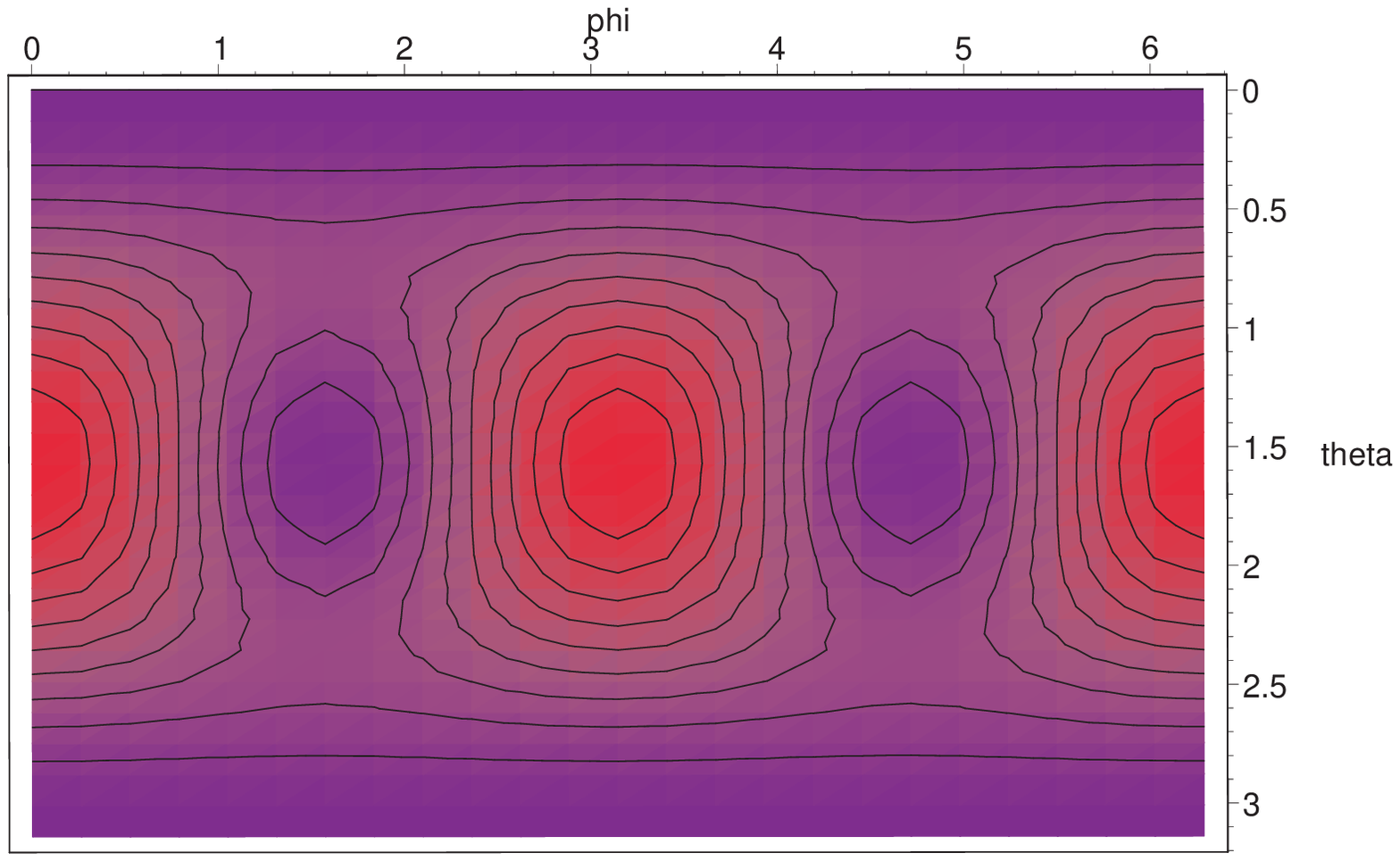}
\begin{minipage}{4.5cm}
$ \epsilon_{1s}$
\end{minipage}
\end{minipage}~~\begin{minipage}{4.5cm}
\includegraphics[width=4.5cm,height=4.5cm]{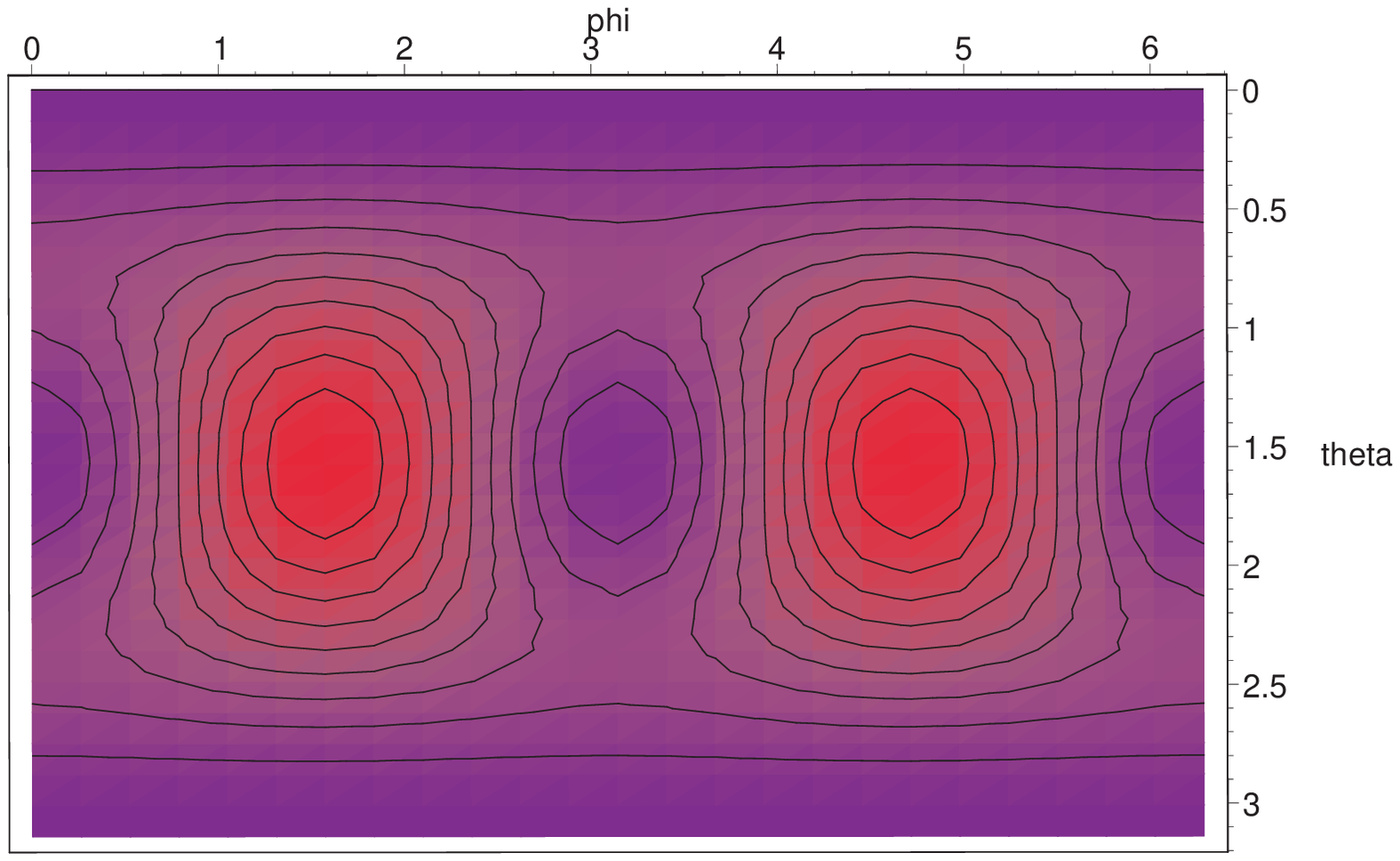}
\begin{minipage}{4.5cm}
$ \epsilon_{1c}$
\end{minipage}
\end{minipage}
\end{center}
\begin{center}
\begin{minipage}{4.5cm}
\includegraphics[width=4.5cm,height=4.5cm]{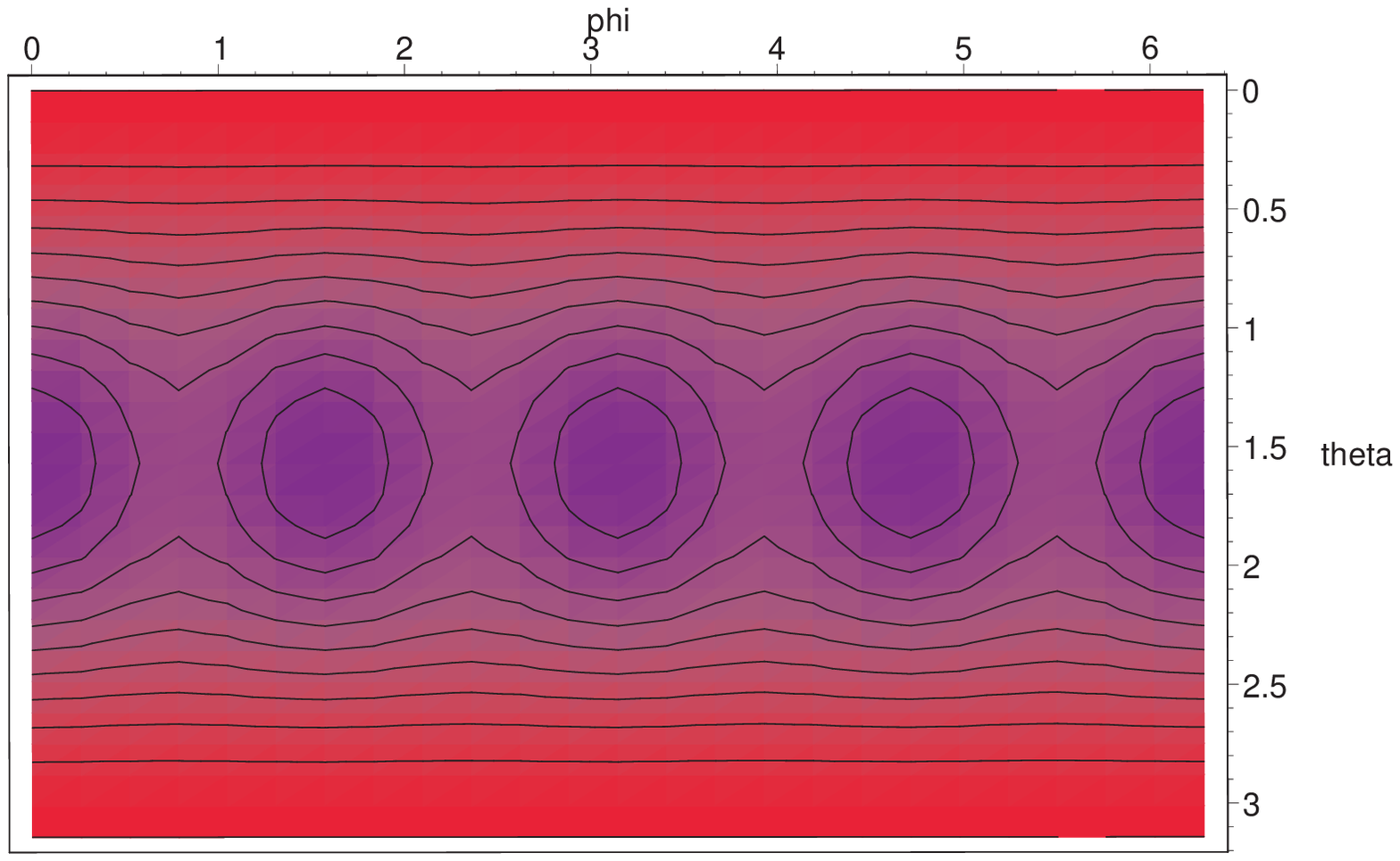}
\begin{minipage}{4.5cm}
$ \epsilon_{2s}$
\end{minipage}
\end{minipage}~~\begin{minipage}{4.5cm}
\includegraphics[width=4.5cm,height=4.5cm]{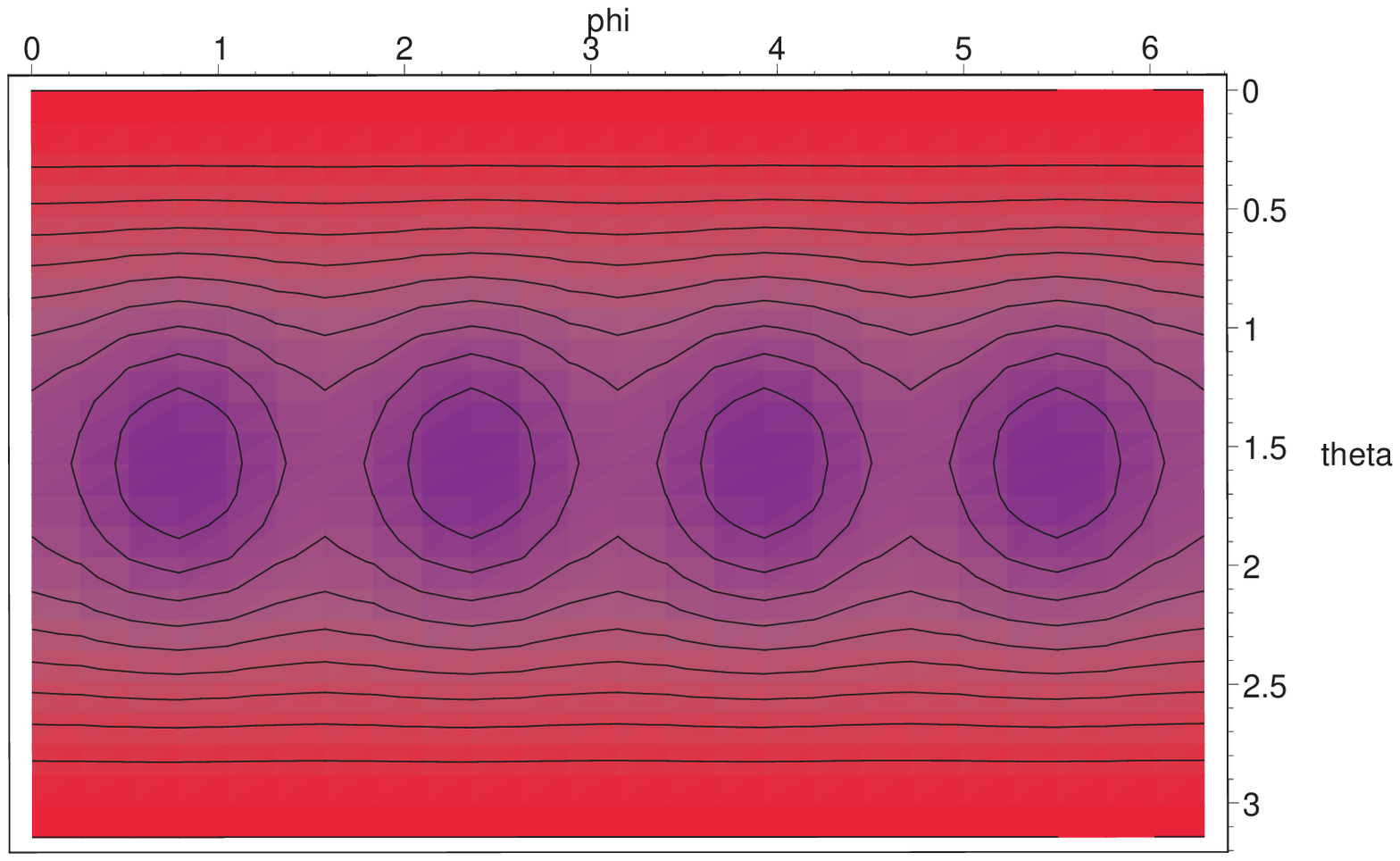}
\begin{minipage}{4.5cm}
$ \epsilon_{2c}$
\end{minipage}
\end{minipage}
\end{center}
\begin{center}
\begin{minipage}{4.5cm}
\includegraphics[width=4.5cm,height=4.5cm]{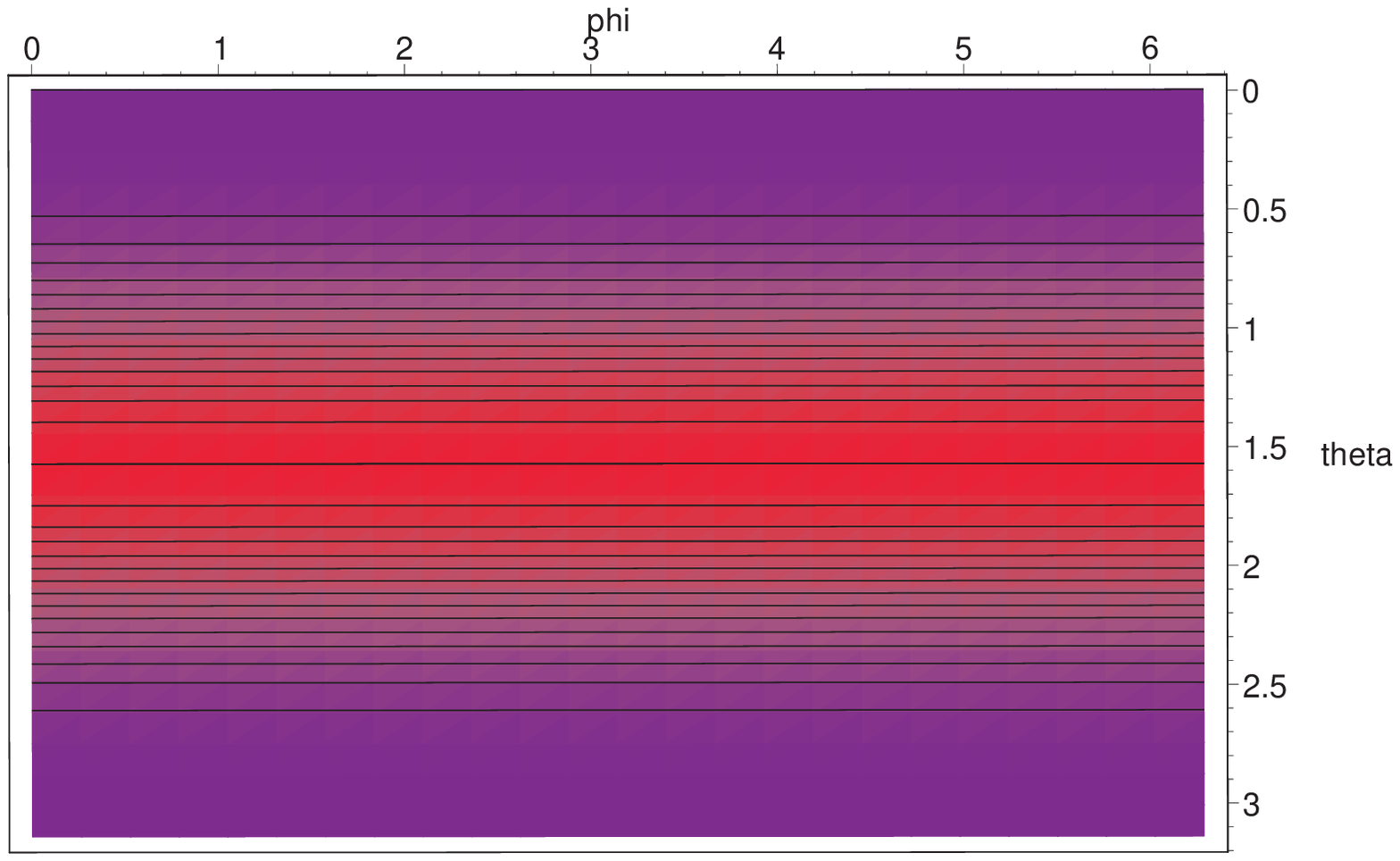}
\begin{minipage}{4.5cm}
$ \epsilon_{0}$
\end{minipage}
\end{minipage}
\caption{Sensitivity of the five fundamental modes as a function of the GW arrival direction $(\theta,\phi).$ The ratio $\epsilon_{\alpha}$ is zero where the graph is blue, and maximal ($\frac{3}{4}$) where the graph is red (Color online).} 
\end{center}
\end{figure}

\begin{figure}\label{fig2}
\begin{center}
\begin{minipage}{4.5cm}
\includegraphics[width=4.5cm,height=4.5cm]{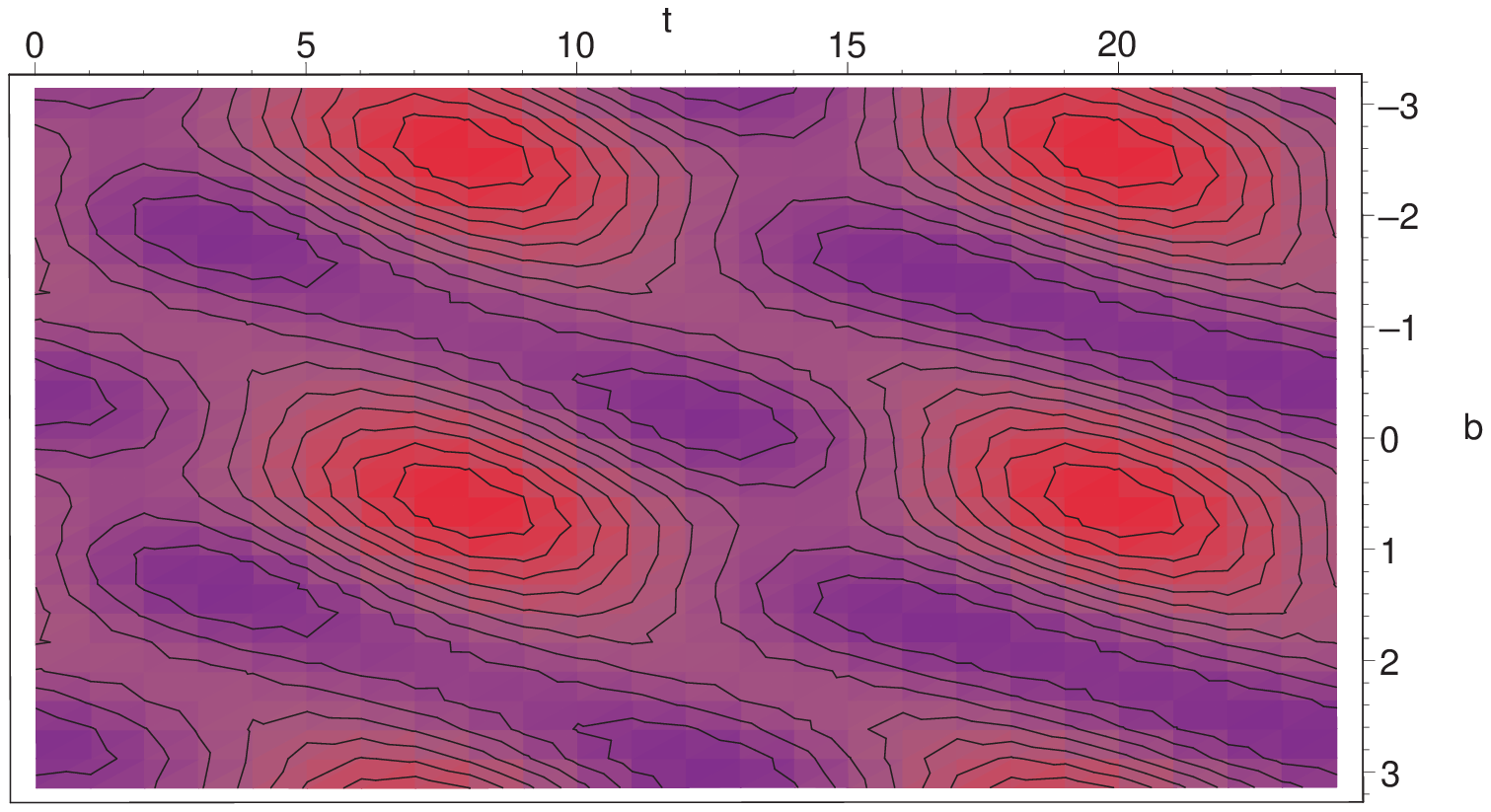}
\begin{minipage}{4.5cm}
\begin{center}
$\epsilon_{1s}$
\end{center}
\end{minipage}
\end{minipage}~~\begin{minipage}{4.5cm}
\includegraphics[width=4.5cm,height=4.5cm]{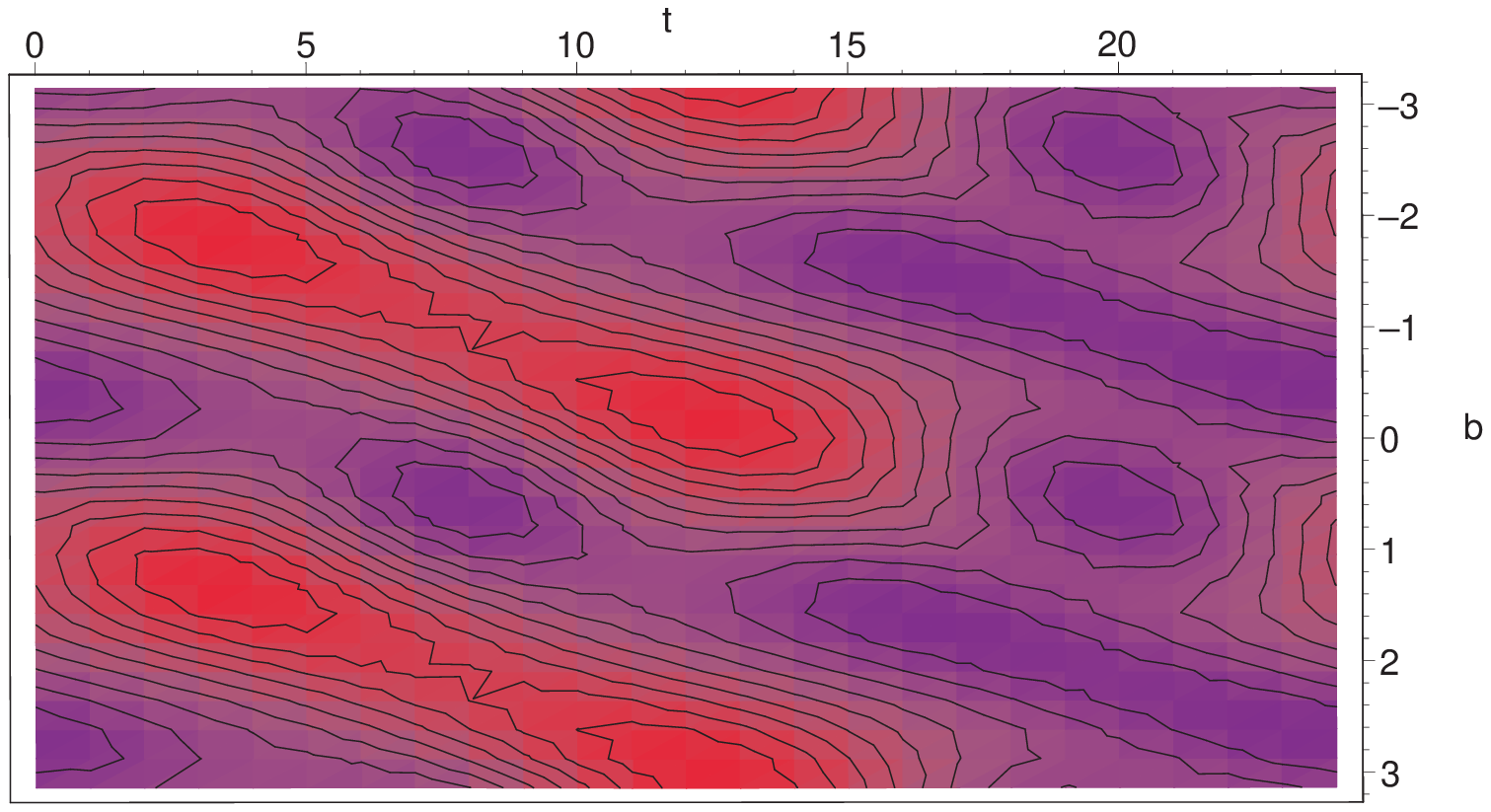}
\begin{minipage}{4.5cm}
\begin{center}
$\epsilon_{1c}$
\end{center}
\end{minipage}
\end{minipage}
\end{center}
\begin{center}
\begin{minipage}{4.5cm}
\includegraphics[width=4.5cm,height=4.5cm]{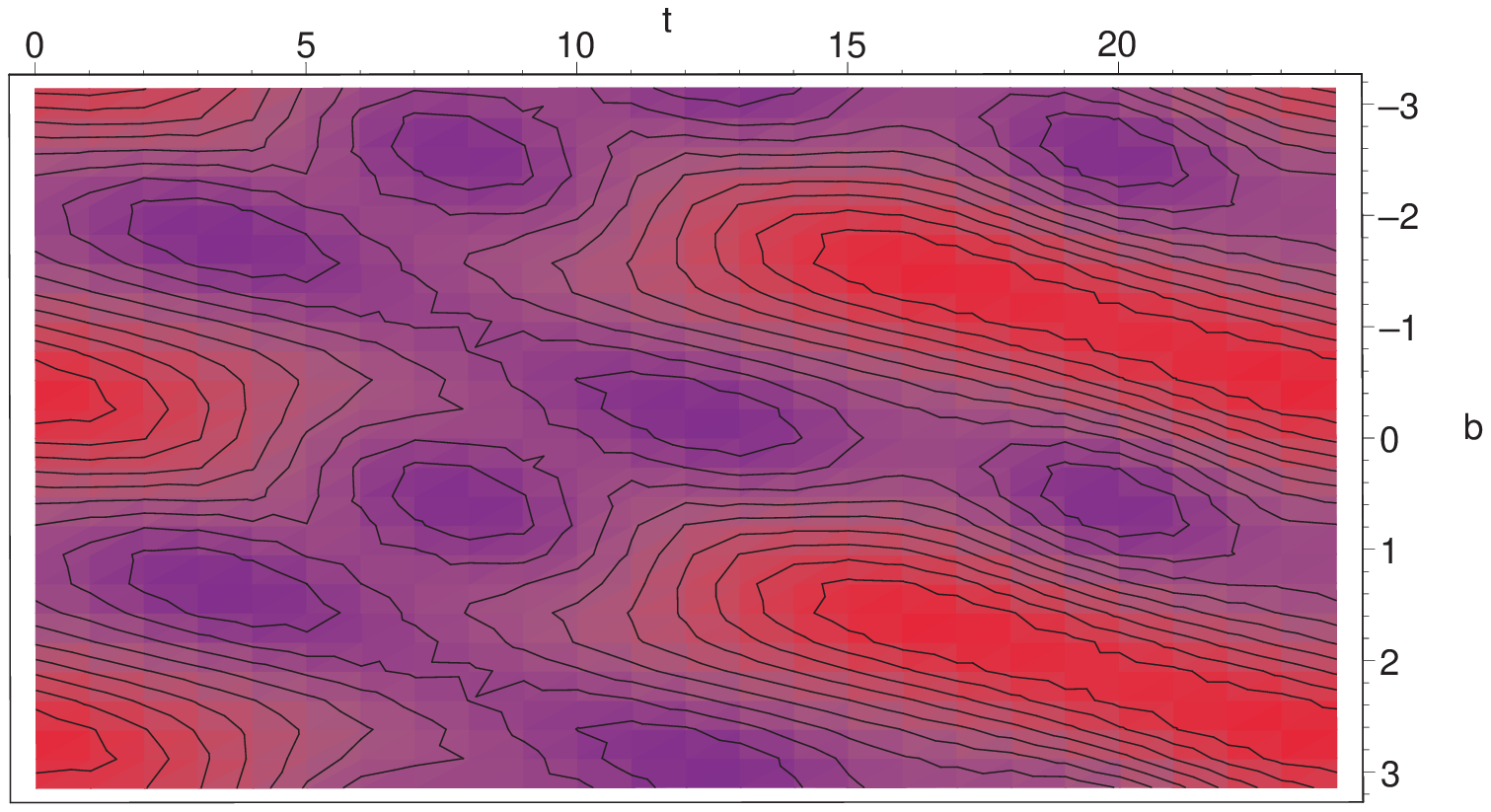}
\begin{minipage}{4.5cm}
\begin{center}
$\epsilon_{2s}$
\end{center}
\end{minipage}
\end{minipage}~~\begin{minipage}{4.5cm}
\includegraphics[width=4.5cm,height=4.5cm]{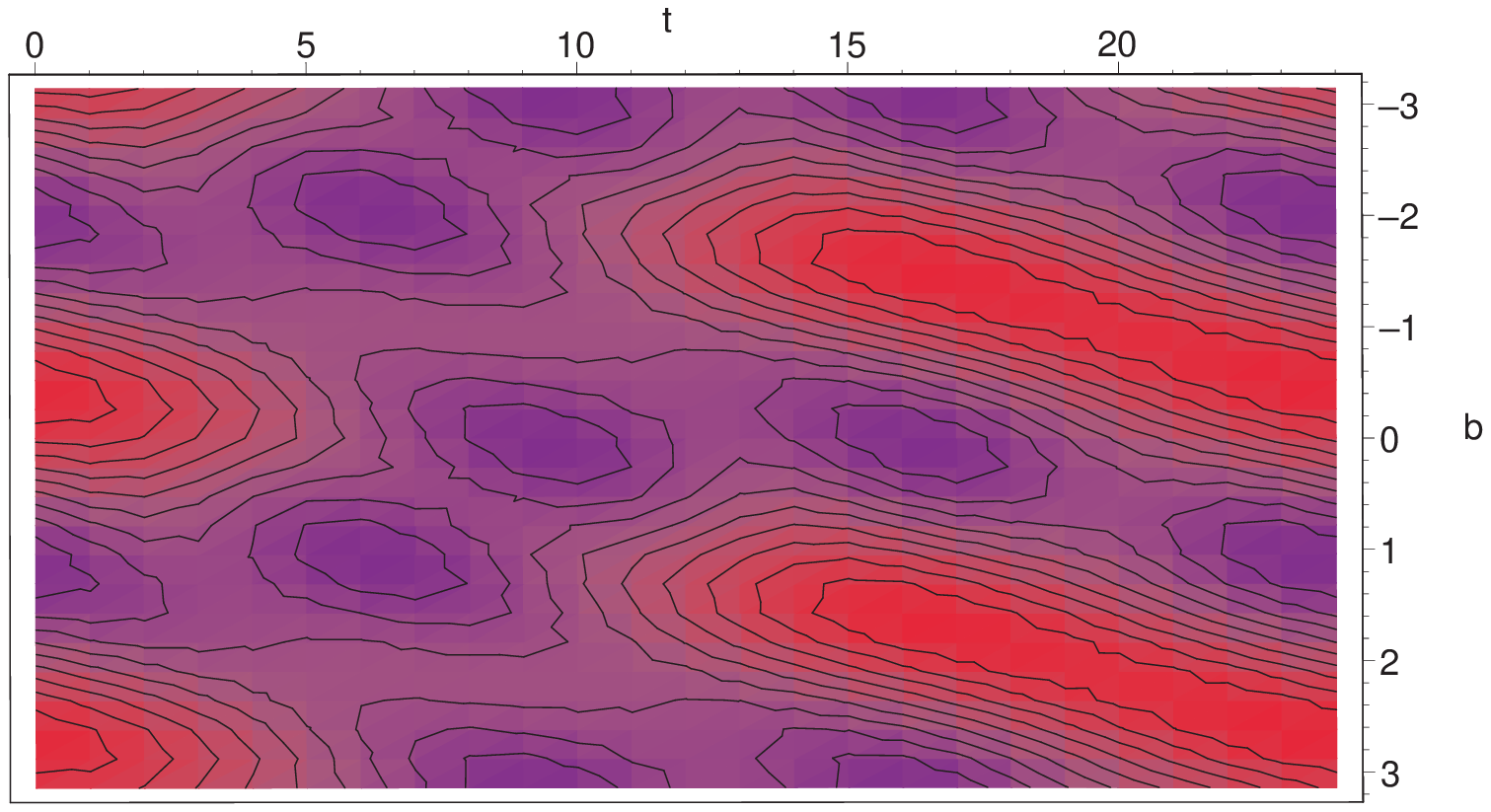}
\begin{minipage}{4.5cm}
\begin{center}
$\epsilon_{2c}$
\end{center}
\end{minipage}
\end{minipage}
\end{center}
\begin{center}
\begin{minipage}{5.5cm}
\includegraphics[width=4.5cm,height=4.5cm]{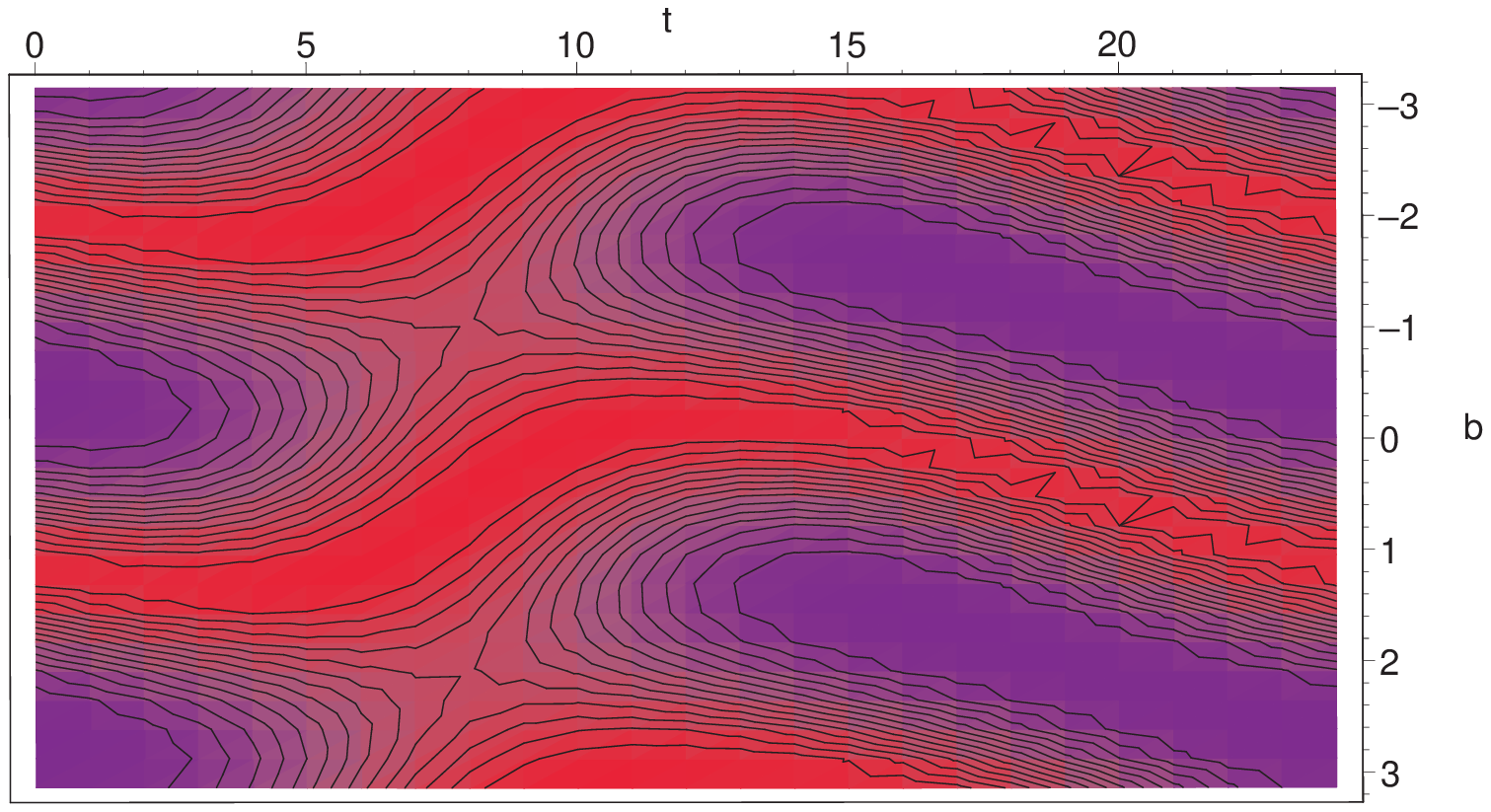}
\begin{minipage}{4.5cm}
\begin{center}
$\epsilon_{0}$
\end{center}
\end{minipage}
\end{minipage}
\end{center}
\caption{ Sensitivity of the five fundamental modes as a function of the GW source position in the galactic plane, $(b),$ and of sidereal time in sidereal hours, ($t$), for a sphere near Leiden, NL (latitude $l=52.16^{\circ}N$, longitude $L=4.45^{\circ}E$). As in Figure 1, red zones indicate maximal values (Color online). }
\end{figure}

\begin{figure}\label{fig3}
\begin{center}
\begin{minipage}{5.5cm}
\includegraphics[width=5.5cm,height=5.5cm]{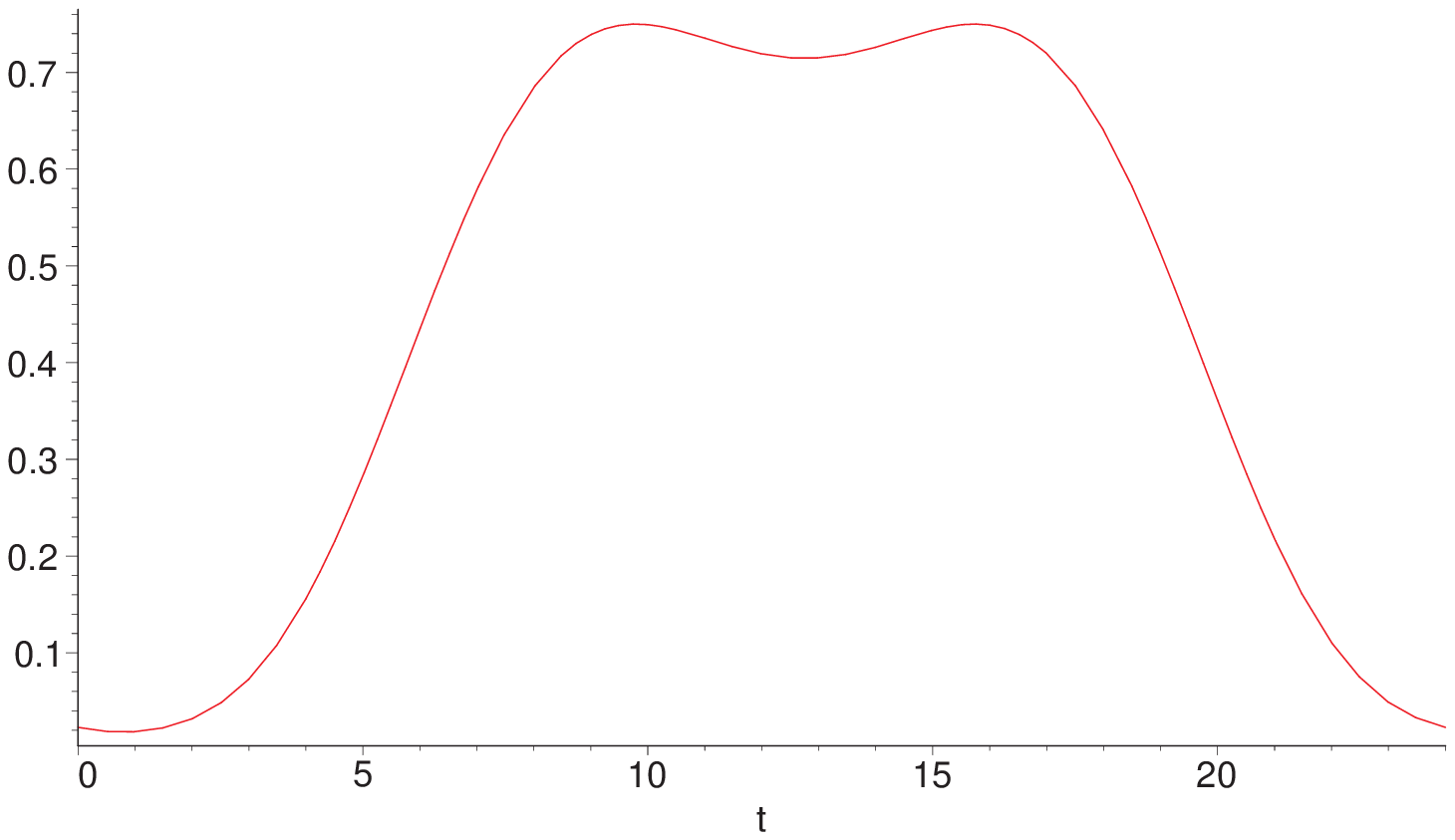}
\end{minipage}
\end{center}
\caption{$\epsilon_{0}$ of Figure 2 as a function of $t$ in sidereal hours for b=0 (source near the galactic center).}
\end{figure}

The total sensitivity is the sum of contributions coming from the five modes $\alpha$ and the sensitivity of each mode depends on the orientation of the source with respect to the chosen frame.
This means that for different angular positions of the source with respect to the detector reference frame, there will be different distributions among the five modes of the total energy transfered from the GW to the sphere.
In contrast, if the source is fixed in space, and the sphere moves because of the earth's rotation, the energy ratio stored in each of the five modes will be a periodic function of the arrival time, with period one sidereal day, while the sum of the five contributions will be time-independent.

The expected sources for GW resonant detectors are traditionally distinguished into: 1) burst of relatively great intensity, given by cataclysmic events as binary cohoaleshing, or supernovae explosions; 2) sources which emit GW bursts occasionally but repeatedly in time and with peculiar statistic features, the so called "GW-bursters" \cite{paper04};
 3) periodic sources, of longer duration and smaller spread in frequency, as pulsars or inspiraling binaries. \\ 
Such astrophysical objets are present in galaxies, and as the energy hitting a detector decreases as the distance of the source squared \cite{Flox}, it is then reasonable to expect a GW signal coming from our galaxy to be more easily detected. 

Once the position of the source is known in galactic coordinates $(b,l_g),$ a set of space rotations gives the position of the source in the detector frame $(\hat{x}_d,\hat{y}_d,\hat{z}_d).$ These rotations are derived and explained in detail in the appendix of \cite{Flox}. 
For a given $(b,l_g)$ one can then easily find the precise position $(\theta(t),\phi(t))$ of the source with respect to the detector frame, with sidereal time $t$ varying from 0 to 24.

As example, figure 2 displays the energy rates $\epsilon_{\alpha}$ stored in the five modes $\alpha$ of a sphere near Leiden, Holland ( latitude $l=52.16^{\circ}N$ , longitude $L=4.45^{\circ}E$, as functions of Greenwich sidereal time (in sidereal hours) and of galactic longitude $b$ for a randomly polarized source lying in the galactic plane ($l_g=0$), where stars population is more important (and then their remnants population also).
As before, red zones indicate maximal values.\\
Finally, the example of a section (at $b=0$, i.e. for a source near the GC) for the $\alpha=0$ mode is represented in Figure 3.
These graphics can be specially interesting for sources of type 2 and 3 mentioned above, for which a signal from the same location persists in time.

\section{ The spherical GW detector with a set of resonators on the surface}

Until now we have looked at the spherical detector as a unique object interacting with GW. Nevertheless, because of the smallness of $h$, the magnitude displacement of the detector as a result of this interaction is tiny, of the order of $h L$, $L$ being the detector size.

For this reason small resonators of equal mass $m_r<<M$ and elastic constant $k_r$ are put on the surface of the GW resonant detector. The small resonators are tuned to the fundamental frequency of the detector, in order to amplify displacements of its surface at that chosen frequency, which for the sphere turns out to be $\omega_{1,2}$. In this way, frequencies other than the fundamental mode and nearby frequencies will not be amplified by the system.

\subsection{ The general case of $K$ resonators}

Imagine having $K$ such resonators on the surface of the sphere. We are interested in what happens when a GW interacts with such a system, in the ideal case of high SNR (i.e. where noises are negligible).\\
In line with previous work, in deriving the equations of motion for such a system we will use the formalism and notations of \cite{M&J}, where the problem has already elegantly been posed, and a proposal for a six-detector configuration, the TIGA configuration, is presented.    

The equation of motion (\ref{eom}) for small displacements now becomes

\be\label{eomres}
\rho\frac{\pa^2\vu(\vr,t)}{\pa t^2}-\mu\nabla^2\vu(\vr,t)-(\la+\mu)\vec{\na}(\vec{\na}\cdot\vu(\vr,t))=\sum_{\eta}\vec{f}^{(\eta)}(\vr)g^{(\eta)}(t)
\ee  
where $\eta$ is a parameter that takes the values ${0,\,1c,\,1s,\,2c,\,2s,\,1,\,2,\,....,\,K}={\{\alpha,\,i\}}.$
It is clear that the first components of this force density, 
\be
\vec{F}_{GW}=\sum_{\al}\vec{f}^{\al}(\vr)g^{\al}(t)=\sum_{\al}\vec{f}^{\al}(\vr)h^{\al}(t),
\ee
 are the usual ones (see Eq.(\ref{fg})) attributable to GW, while
\be\label{fi}
\vec{F}_{res}=\sum_{i=1}^{K}\vec{f}^{(i)}(\vr) g^{(i)}(t)
\ee are new contributions given by the elastic terms of resonators.

Let us focus now on one resonator, which we assume to have only radial motion, linked with a tuned spring to the surface of the sphere, say at the position $\vec{R}_k$. We define the scalar-time dependent quantity 
\be\label{z}
z_i(t)=\vec{u}(\vec{R}_i,t)\cdot\hat{r}_i
\ee 
 as the radial projection of the surface displacement at this point. \\
We also define $q_i(t)$ to be the relative distance between the resonator and the sphere surface. Note that $q_i(t)$ is not an inertial coordinate, but $z_i(t)+q_i(t)$ is. Then the equation of motion for this last quantity is
\bea
\label{z+qeom}
m_r(\ddot{q}_i(t)+\ddot{z}_i(t))=-k_r q_i(t)+F_i^{GW}(t),
\eea
where $F_i^{GW}$ is the radial component of the GW force density acting on the resonator, which will be neglected as we know that $m_r<<M$.

Since
\be
g^{i}(t)=q_i(t)\quad,\quad\vec{f}^{i}(\vr)=k_r\de^3(\vr-\vec{R}_i)\hat{R}_i,
\ee
the force density component (\ref{fi}) can be written: 
\be\label{Fi}
\vec{F}_{res}=\sum_i k_r q_i(t)\de^3(\vr-\vec{R}_i)\hat{R}_i.
\ee

Generalizing the same method used above (in particular, see Eqs.(\ref{SolGenerale}), (\ref{fgfrak}) and (\ref{B}) ), the solution of Eq.(\ref{eomres}) can be expressed as:
\bea
\label{Balpha}
\vu(\vr,t)&=&\sum_{\eta}\sum_{n,l,\alpha}\omega_{n,l,\alpha}^{-1}\mathfrak{f}_{n,l,\alpha}^{\eta}\mathfrak{g}_{n,l,\alpha}^{\eta}(t)\vec{\Phi}_{n,l,\alpha}(\vr)\nonumber\\
&=&\sum_{n,l,\alpha}B_{n,l,\alpha}(t)\vec{\Phi}_{n,l,\alpha}(\vr)
\eea
 with 
 \bea
\mathfrak{f}_{n,l,\alpha}^{\eta}&=&\frac{1}{M}\int_{Sphere}\vec{\Phi}_{n,l,\alpha}(\vr)\cdot\vec{f}^{\eta}(\vr)d^3r\nonumber\\
\mathfrak{g}_{n,l,\alpha}^{\eta}(t)&=&\int_0^t g^{\eta}(t')\sin(\omega_{n,l,\alpha}(t-t'))dt'.
  \eea
 The last equality of Eq.(\ref{Balpha}) defines
 \be
 B_{n,l,\alpha}(t)=\sum_{\eta}\omega_{n,l,\alpha}^{-1}\mathfrak{f}_{n,l,\alpha}^{\eta}\mathfrak{g}_{n,l,\alpha}^{\eta}(t).
 \ee
 
 In our case we get
 \be
 \mathfrak{f}^{\eta}_{n,l,\alpha}=\left\{\begin{array}{ll}
 a_{n,l}\de_{\alpha,\alpha'} & \textrm{if $\eta=\alpha'$}\\
 \frac{k_r}{M}A_{n,l}(R)Y_{n,l,\alpha}(\theta_i,\phi_i) & \textrm{if $\eta=i$ .}
 \end{array} \right.
 \ee

It is then straightforward to derive the equations of motion for the sphere modes $B_{n,l,\alpha}(t),$ by inserting Eq.(\ref{Balpha}) into Eq.(\ref{eomres}) :
\bea
\label{Beomot}
&&\ddot{B}_{n,l,\alpha}(t)+\omega_{n,l,\alpha}^2B_{n,l,\alpha}(t)=\sum_{\eta}\mathfrak{f}_{n,l,\alpha}^{\eta}g^{\eta}(t)\nonumber\\
&&=a_{n,l}\ddot{h}^{\al}(t)+\sum_{i=1}^K\frac{k_r}{M}A_{n,l}(R)Y_{n,l,\alpha}(\theta_i,\phi_i)q_i(t).\nonumber\\
\eea

When noises are negligible, the motion is dominated by quadrupole modes ($l=2,\,\alpha=0,\,1c,\,1s,\,2c,\,2s$), the only ones interacting with GW. We limit ourselves to the $n=1$ modes, for which the cross-section is the most significant \cite{Coccy}. Note that for $l=2$, $a_{n,2}\equiv a_n$ of Eq.(\ref{a_n}).\\ From now on, we simplify the notation, writing
\bea
B_{1,2,\alpha}(t)  =  B_{\alpha}(t)\quad,\quad a_{1,2} & = & a\quad,\quad A_{1,2}(R)=A \nonumber\\ Y_{1,2,\alpha}(\theta_i,\phi_i) & = & Y_{\al}(\theta_i,\phi_i)=Y_{\alpha i}.
\eea
 This last equality defines the $5\times K$ matrix $\Y.$

Multiplying for the sphere mass $M$, and calling the five equal quantities $M\omega_{1,2,\alpha}^2\equiv M\omega_s^2=k_s,$ Eq.(\ref{Beomot}) can be  rewritten
\be\label{Beom'}
M\ddot{B}_{\alpha}(t)+k_sB_{\alpha}(t)=Ma\ddot{h}^{\al}(t)-\sum_{i}k_rAY_{\a i}q_i(t).
\ee
Defining now the vectors 
\bea
&&\{q_i(t)\}=\q(t)\; ,\; \{Ma\ddot{h}^{\al}(t)\}=\{\ddot{H}^{\alpha}(t)\}=\ddot{\H}(t) \;,\nonumber\\ && \{B_{\a}(t)\}=\b(t)\; ,
\eea 
we can again simplify the expression for these equations:
\be\label{matrix1}
M\I_5\ddot{\b}(t)+k_s\I_5\b(t)-k_r A\Y\q(t)=\ddot{\mathbf{H}}(t).
\ee
It is evident that $\b$ and $\h$ are 5-component vectors, $\q$ is a $K$-component vector and $\I_5$ the identity matrix $5\times 5.$ Thus Eq.(\ref{matrix1}) is a system of 5 equations.\\
In the same way we can rewrite Eq.(\ref{z+qeom}) in matrix form. To do that, we need to express $z_i$ as a function of $B_{\al}$. Using the approximation that only the $n=1$ modes of the sphere are excited at high SNR, from Eq.(\ref{z}) we have:
\be
z_i(t)=\vec{u}(\vec{R}_i,t)\cdot\hat{r}_i\cong\sum_{n=1,l=2,\a}B_{\a}(t)AY_{\a i}.
\ee 
Defining the vector $\{z_i\}(t)=\mathbf{z}(t)$, 
\be\label{vecz}
\mathbf{z}(t)=A\Y^T\b(t).
\ee
Then Eq.(\ref{z+qeom}) becomes
\be\label{matrix2}
m_r(\ddot{\q}(t)+A\Y^T\ddot{\b}(t))+k_r\q(t)=\mathbf{0}.
\ee
The $K$ equations for $\q$ (\ref{matrix2}) and the 5 equations for $\b$ (\ref{matrix1}) can be rewritten as a system of $K+5$ equations
\bea
\label{star}
&&\begin{displaymath}
\begin{array}{cc}
M\I_5 & \0 \\
m_rA\Y^T & m_r\I_K
\end{array}
\end{displaymath}
\cdot
\begin{displaymath}
\begin{array}{c}
\ddot{\b} \\
\ddot{\q}
\end{array}
\end{displaymath}
(t)\nonumber\\&+&
\begin{displaymath}
\begin{array}{cc}
k_s\I_5 & -k_rA\Y \\
\0 & k_r\I_K
\end{array}
\end{displaymath}
\cdot
\begin{displaymath}
\begin{array}{c}
\b \\
\q
\end{array}
\end{displaymath}
(t)=\begin{displaymath}
\begin{array}{c}
\ddot{\H}(t) \\
\0
\end{array}
\end{displaymath}.
\eea

These are the equations of motion for a system composed of a homogeneous sphere with $K$ resonators moving radially on its surface, all interacting with a GW. 

In the remainder of this paper, we will solve these equations once a configuration for the positions of the resonators is given. We first look at what happens in the case with only one resonator ($k=1$), in order to test the equations by looking at a simple example. Then we move on to the TIGA configuration ($K=6$) proposed by Merkowitz and Johnson in \cite{M&J}.\\

\subsection{The case of one resonator at $\theta=0$}

In this particular case we choose to put only one resonator on the surface of the sphere. Obviously each position is equivalent, but for simplicity let us put the resonator at the north pole of our coordinates system (or, equivalently, at the south pole because of the symmetry of the problem).

In this case, the only nonvanishing component of the $5\times 1$ matrix $\Y$ is $Y_{01}=\sqrt{\frac{5}{4\pi}}=y.$ Then Eq.(\ref{star}) can be written: 
\bea
\begin{displaymath}
\begin{array}{cccccc}
M & 0 & 0 & 0 & 0 & 0 \\
0 & M & 0 & 0 & 0 & 0 \\
0 & 0 & M & 0 & 0 & 0 \\
0 & 0 & 0 & M & 0 & 0 \\
0 & 0 & 0 & 0 & M & 0 \\
myA & 0 & 0 & 0 & 0 & m\\
\end{array}
\end{displaymath}
&\cdot&
\begin{displaymath}
\begin{array}{c}
\ddot{B_0} \\
\ddot{B}_{1c} \\
\ddot{B}_{1s} \\
\ddot{B}_{2c} \\
\ddot{B}_{2s} \\
\ddot{q}\\
\end{array}
\end{displaymath}\nonumber \\
+
\begin{displaymath}
\begin{array}{cccccc}
k_s & 0 & 0 & 0 & 0 & -k_rAy \\
0 & k_s & 0 & 0 & 0 & 0 \\
0 & 0 & k_s & 0 & 0 & 0 \\
0 & 0 & 0 & k_s & 0 & 0 \\
0 & 0 & 0 & 0 & k_s & 0 \\
0 & 0 & 0 & 0 & 0 & k_r\\
\end{array}
\end{displaymath}
&\cdot&
\begin{displaymath}
\begin{array}{c}
B^0 \\
{B}^{1c} \\
{B}^{1s} \\
{B}^{2c} \\
{B}^{2s} \\
{q}\\
\end{array}
\end{displaymath}
=
\begin{displaymath}
\begin{array}{c}
\ddot{H}^0 \\
\ddot{H}^{1c} \\
\ddot{H}^{1s} \\
\ddot{H}^{2c} \\
\ddot{H}^{2s}\\
{0}\\
\end{array}
\end{displaymath}.\label{star1}\nonumber\\
\eea
 
It is clear from Eq.(\ref{star1}) that the only mode of the sphere which couples with our resonator is $B_0$, while all the others behave as if the resonator was not present. Their equations of motion are the same as given above in Eq.(\ref{Beomsf}).\\
So we can reduce our system of equations to a system of 2 coupled equations with 2 unknowns,
taking only the first and the last lines of (\ref{star1}):
\bea\label{eomqB}
\ddot{B}_0(t)+\omega_0^2B_0&=&\omega_0^2Ay\epsilon q+a\ddot{h}^0(t)\nonumber\\
\ddot{q}(t)+\omega_0^2q(t)&=&-Ay\ddot{B}_0(t).
\eea

In Eq.(\ref{eomqB}) we consider that the resonator frequency is perfectly tuned with the fundamental frequency of the sphere, i.e. that $$\frac{k_s}{M}=\frac{k_r}{m}=\omega_0^2,$$ and we call $\epsilon $ the (small) ratio between the resonator mass and the sphere mass.
 
In order to solve Eq.(\ref{eomqB}) we move to Fourier coordinates, giving:
\bea\label{qBzero}
\tilde{q}(\omega)&=&\frac{-\omega^4aAy\tilde{h}^0(\omega)}{(\omega_0^2-\omega^2)^2-\omega^2\omega_0^2A^2y^2\epsilon}\nonumber\\
\tilde{B}_0(\omega)&=&a\omega^2\tilde{h}^0(\omega)\frac{\omega^2-\omega_0^2}{(\omega_0^2-\omega^2)^2-\omega_0^2\omega^2A^2y^2\epsilon}.
\eea

As expected, the resonant frequencies $\omega_{\pm}$ for both the coupled quantities are the same, and they are the splitting of the original frequency $\omega_0$.\\
They can be found setting the denominators of Eq.(\ref{qBzero}) equal to zero:
\bea
\omega_{\pm}^2&=&\omega_0^2
\left(1\pm Ay\sqrt{\epsilon\left(1+\frac{A^2y^2\epsilon}{4}\right)}+\frac{A^2y^2\epsilon}{2}\right)=\nonumber\\
&=&\omega_0^2\left(1\pm\sqrt{\epsilon A^2y^2}+  \frac{A^2y^2\epsilon}{2}\pm\frac{(A^2y^2\epsilon)^{3/2}}{8}+O(\epsilon^{2})\right).\nonumber\\
\eea 

At the frequency $\omega_0$ we have $\tilde{B}_0(\omega_0)=0,$ and  $\tilde{q}(\omega_0)=\frac{a}{Ay\epsilon}\tilde{h}^0(\omega_0)$  is proportional to the ratio between the sphere mass and the resonator mass, so we will have the greatest oscillations if we make the resonator mass as small as possible.

Note that here we neglected the damping of both the coupled oscillators. This is because we are interested in burst signals, where the time-scale for damping an oscillation of the system is much longer than the time duration of a signal. 
Taking account of damping means changing Eq.(\ref{qBzero}) to cancel divergences which arise at $\omega=\omega_{\pm}$ and make them smoother. In particular, $\tilde{B}_0(\omega_0)$ will no longer vanish.
Computations adding damping terms are analog and easily done. Figure 4 shows a plot of $|\tilde{q}(\omega)|^2$ with a sphere damping of $10^{-4}$, a resonator damping of $10^{-2}$, $\omega_0=10^3Hz$, $\epsilon=10^{-6},$ and $\tilde{h}^0(\omega)=10^{-19}Hz^{-1}.$  
\begin{figure}\label{fig4}
\begin{center}
\includegraphics[width=5.5cm,height=5.5cm]{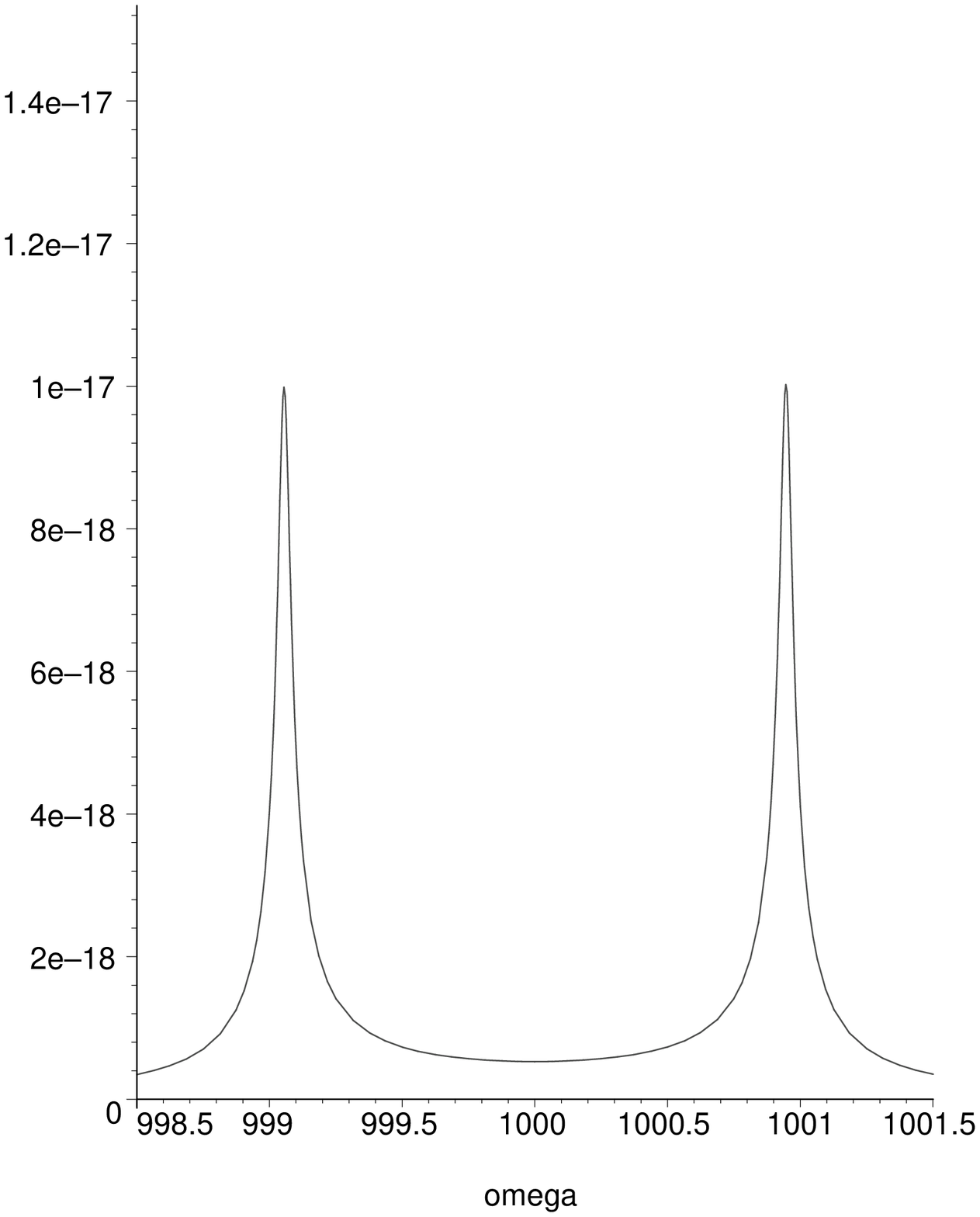}
\end{center}
\caption{$|\tilde{q}(\omega)|^2$ with a sphere damping of $10^{-4}$, a resonator damping of $10^{-2}$, $\omega_0=10^3Hz$, $\epsilon=10^{-6},$ and $\tilde{h}^0(\omega)=10^{-19}Hz^{-1}.$ }
\end{figure}

The quantity we are really interested in is the one we can measure, $\tilde{q}(\omega),$ or its squared absolute value. Rewriting $\tilde{h}^0$ to show its dependence on $\tilde{h}_+$, $\tilde{h}_{\times}$ and on the wave direction (see Eq.(\ref{g})) we get:
\bea
|\tilde{q}(\omega)|^2&=&\frac{\omega^8a^2A^2}{4\left[(\omega_0^2-\omega^2)^2-\omega_0^2\omega^2A^2y^2\epsilon\right]^2}\sin^4\theta|\tilde{h}_+(\omega)|^2\nonumber\\
&=&\beta(\omega)\sin^4\theta|\tilde{h}_+(\omega)|^2,
\eea
or, assuming that the GW is unpolarized, and averaging over polarizations,
\bea
\frac{|\tilde{q}(\omega)|^2}{\beta(\omega)(|\tilde{e}_+|^2+|\tilde{e}_{\times}|^2)/2}=\sin^4(\theta),
\eea
with 
\be\label{beta}
\beta(\omega)=\frac{\omega^8a^2A^2}{4(\omega^2-\omega_+^2)(\omega^2-\omega_-^2)}.
\ee
This turns out to be the same angular sensitivity obtained for the cylindrical bar \cite{Thorne} , for which one finds that sensitivity is optimal for a plane orthogonal to the bar and drops with a factor $\sin^4\theta$ too.\\
Having only one transducer on the sphere surface is a simple example showing that the way resonators are positioned can result in the loss of the main initial properties given by the spherical symmetry of the detector. In order to keep these properties as far as possible, we have to put more resonators on the sphere in appropriate positions.

\subsection{The TIGA configuration}

With the hope that, in future, we will be able to place resonators moving radially on the sphere surface, and that we will be able to measure experimentally the displacements $\q$ of these masses, and assuming that all other interactions with the sphere-resonators system are negligible\footnote{In particular, we consider the case where another amplification device, such as SQID, is NOT tuned to the frequency of resonance of the sphere.} with the exception of gravitational waves, in this section we find the response of $\tilde{\q}(\omega)$ in the TIGA configuration \cite{M&J} to a gravitational wave coming from a generic direction $(\theta,\phi).$
In order to do so, we have to solve the equations of motion (\ref{star}) in Fourier space.

In the TIGA configuration there are a total of six resonators: three of them have azimuth angle $\theta_1=\theta_2=\theta_3=\theta_A$ and $\phi_1=0,\,\phi_2=\frac{2\pi}{3},\,\phi_3=\frac{4\pi}{3}$ and the other three have $\theta_4=\theta_5=\theta_6=\theta_B$ and $\phi_4=\frac{\pi}{3},\,\phi_5=\pi,\,\phi_6=\frac{5\pi}{3}$. The angles $\theta_A$ and $\theta_B$ are between $0$ and $\pi/2$ and they are solutions of the equation
$$45\cos^4\theta-30\cos^2\theta+1=0.$$    
With this information, the $\Y$ matrix for this configuration is easily found:

\begin{widetext}
\bea
\begin{displaymath}
\begin{array}{cccccc}
\frac{1}{\sqrt{4\pi}} & \frac{1}{\sqrt{4\pi}} & \frac{1}{\sqrt{4\pi}} & -\frac{1}{\sqrt{4\pi}} & -\frac{1}{\sqrt{4\pi}} & -\frac{1}{\sqrt{4\pi}} \\
\\
\sqrt{\frac{3+\sqrt{5}}{6\pi}} & -\sqrt{\frac{3+\sqrt{5}}{24\pi}} & -\sqrt{\frac{3+\sqrt{5}}{24\pi}} & \sqrt{\frac{3-\sqrt{5}}{24\pi}} & -\sqrt{\frac{3-\sqrt{5}}{6\pi}} & \sqrt{\frac{3-\sqrt{5}}{24\pi}} \\
\\
0 & \sqrt{\frac{3+\sqrt{5}}{8\pi}} & -\sqrt{\frac{3+\sqrt{5}}{8\pi}} & \sqrt{\frac{3-\sqrt{5}}{8\pi}} & 0 & -\sqrt{\frac{3-\sqrt{5}}{8\pi}} \\
\\
\sqrt{\frac{3-\sqrt{5}}{6\pi}} & -\sqrt{\frac{3-\sqrt{5}}{24\pi}} & -\sqrt{\frac{3-\sqrt{5}}{24\pi}} & -\sqrt{\frac{3+\sqrt{5}}{24\pi}} & \sqrt{\frac{3+\sqrt{5}}{6\pi}} & -\sqrt{\frac{3+\sqrt{5}}{24\pi}} \\
\\
0 & -\sqrt{\frac{3-\sqrt{5}}{8\pi}} & \sqrt{\frac{3-\sqrt{5}}{8\pi}} & \sqrt{\frac{3+\sqrt{5}}{8\pi}} & 0 & -\sqrt{\frac{3+\sqrt{5}}{8\pi}} \\
\end{array}
\end{displaymath}
\eea
\end{widetext}
It is important to recall a very usefull property of this matrix:  $\Y\Y^T=\frac{3}{2\pi}\I_5$  is proportional to the identity $5\times 5$ matrix. Nevertheless, the inverse is not true: $\Y^T\Y$ has all diagonal values equal to $\frac{5}{4\pi}$ and all non diagonal values equal to $-\frac{1}{4\pi}.$ It is easily seen that sums over lines or columns are always zero. Actually, $\Y^T\Y$ is a $6\times 6$ matrix obtained from two $5\times 6$ matrices, so it is clear that it cannot be invertible.

Using this property it is straightforward to solve the equations for the so-called $\it{mode- channels}$ \cite{Merkothese} 
\be\label{p}
\tilde{\p}(\omega)=\Y\tilde{\q}(\omega),
\ee

\be
\label{modch}
\tilde{p}_{\alpha}(\omega)=\frac{-Aa\omega^4\tilde{h}^{\alpha}(\omega)}{\frac{2\pi}{3}(\omega_0^2-\omega^2)^2-\epsilon A^2\omega^2\omega_0^2},
\ee
and for $\tilde{\b}(\omega)$
\be
\tilde{B}_{\alpha}(\omega)=\frac{(\omega^2-\omega_0^2)\omega^2 a \tilde{h}^{\alpha}(\omega)}{(\omega_0^2-\omega^2)^2-\frac{3}{2\pi}\epsilon A^2\omega^2\omega_0^2}.
\ee

The denominators vanish for the two split frequencies
\be
\omega^2_{\pm}=\omega_0^2\left[1\pm\frac{3}{4\pi}\sqrt{\frac{8\pi}{3}\epsilon A^2+\frac{\epsilon^2 A^4}{\omega_0^2}}+\frac{3}{4\pi}\epsilon A^2\right],
\ee
so one can rewrite Eq.(\ref{modch}) as
\bea
\tilde{p}_{\alpha}(\omega)&=&-\frac{3}{2\pi}\frac{Aa\omega^4\tilde{h}^{\alpha}(\omega)}{(\omega^2-\omega_+^2)(\omega^2-\omega_+^2)}\nonumber\\
&=&-\left[\frac{\lambda_+}{\omega^2-\omega_+^2}+\frac{\lambda_-}{\omega^2-\omega_-^2}  \right]\frac{3}{2\pi}\omega^2 A\tilde{h}^{\alpha},
\eea
with
\be
\lambda_+=\frac{\omega_+^2}{\omega_+^2-\omega_-^2}\quad,\quad \lambda_-=\frac{\omega_-^2}{\omega_-^2-\omega_+^2}
.\ee

We then have an expression giving a linear combination of resonator displacements as a function of the spherical components of a gravitational wave, or we have each spherical component of $\tilde{h}$ as a function of different combinations of displacements.
 
Nevertheless we are also interested in the inverse relation: we want to write each $\tilde{q}_i$ as a combination of the $\tilde{h}^{\alpha}.$ 
In order to do that we have to express 6 quantities as a function of 5, which means inverting the matrix $\Y,$  a $5\times 6$ matrix. We then need a supplementary condition for $\q$. The condition to be added is true for the TIGA configuration only, in the case of a high SNR. In that case, the sum of all the six displacements is zero, and this can be seen by looking at the only non-trivial solution of $\Y\q=\0.$
This means that we can add a sixth line to the $\Y$ matrix, a line made up of six unities, and a sixth null component to the $\p$ vector in Eq.(\ref{p}). We will call the new $6\times 6$ matrix $\hat{\Y}$ and the new 6-vector $\hat{\p}.$

$\hat{\Y}$ is invertible, and we can easily find $\q$ using $\q=\hat{\Y}^{-1}\hat{\p},$ getting:
\begin{widetext}
\bea
\tilde{q}_1(\omega)&=&\frac{\sqrt{\pi}}{3}\tilde{p}_{0}(\omega)+\frac{\sqrt{3\pi}}{9}(\sqrt{5}+1)\tilde{p}_{1c}(\omega)+\frac{\sqrt{3\pi}}{9}(\sqrt{5}-1)\tilde{p}_{2c}(\omega);\nonumber\\
\tilde{q}_2(\omega)&=&\frac{\sqrt{\pi}}{3}\tilde{p}_{0}(\omega)-\frac{\sqrt{3\pi}}{18}(\sqrt{5}+1)\tilde{p}_{1c}(\omega)+\frac{\sqrt{\pi}}{6}(\sqrt{5}+1)\tilde{p}_{1s}(\omega)-\frac{\sqrt{3\pi}}{18}(\sqrt{5}-1)\tilde{p}_{2c}(\omega)\nonumber\\
&-&\frac{\sqrt{\pi}}{6}(\sqrt{5}-1)\tilde{p}_{2s}(\omega);\nonumber\\
\tilde{q}_3(\omega)&=&\frac{\sqrt{\pi}}{3}\tilde{p}_{0}(\omega)-\frac{\sqrt{3\pi}}{18}(\sqrt{5}+1)\tilde{p}_{1c}(\omega)-\frac{\sqrt{\pi}}{6}(\sqrt{5}+1)\tilde{p}_{1s}(\omega)-\frac{\sqrt{3\pi}}{18}(\sqrt{5}-1)\tilde{p}_{2c}(\omega)\nonumber\\
&+&\frac{\sqrt{\pi}}{6}(\sqrt{5}-1)\tilde{p}_{2s}(\omega);\nonumber\\
\tilde{q}_4(\omega)&=&-\frac{\sqrt{\pi}}{3}\tilde{p}_{0}(\omega)+\frac{\sqrt{3\pi}}{18}(\sqrt{5}-1)\tilde{p}_{1c}(\omega)+\frac{\sqrt{\pi}}{6}(\sqrt{5}-1)\tilde{p}_{1s}(\omega)-\frac{\sqrt{3\pi}}{18}(\sqrt{5}+1)\tilde{p}_{2c}(\omega)\nonumber\\
&+&\frac{\sqrt{\pi}}{6}(\sqrt{5}+1)\tilde{p}_{2s}(\omega);\nonumber\\
\tilde{q}_5(\omega)&=&-\frac{\sqrt{\pi}}{3}\tilde{p}_{0}(\omega)-\frac{\sqrt{3\pi}}{9}(\sqrt{5}-1)\tilde{p}_{1c}(\omega)+\frac{\sqrt{3\pi}}{9}(\sqrt{5}+1)\tilde{p}_{2c}(\omega);\nonumber\\
\tilde{q}_6(\omega)&=&-\frac{\sqrt{\pi}}{3}\tilde{p}_{0}(\omega)+\frac{\sqrt{3\pi}}{18}(\sqrt{5}-1)\tilde{p}_{1c}(\omega)-\frac{\sqrt{\pi}}{6}(\sqrt{5}-1)\tilde{p}_{1s}(\omega)-\frac{\sqrt{3\pi}}{18}(\sqrt{5}+1)\tilde{p}_{2c}(\omega)\nonumber\\
&-&\frac{\sqrt{\pi}}{6}(\sqrt{5}+1)\tilde{p}_{2s}(\omega).
\eea
\end{widetext}
Taking the squared absolute value, and rewriting $\tilde{h}^{\alpha}(\omega)$ as a function of $\tilde{h}_+\,,\,\tilde{h}_{\times}$ and of the direction $(\theta,\phi)$ (see Eq.(\ref{g})) we can find the explicit behavior of the amplitude of the oscillations for each resonator as a function of the GW direction and polarization.

Supposing the source to be randomly polarized, we can replace $\tilde{h}_{+,\times}$ with the expressions in Eq.(\ref{pol}), and average over polarizations (as we have already done for the sphere without resonators). This is equivalent to replacing
\bea
|\tilde{h}_+|^2&\rightarrow &\frac{|\tilde{e}_+|^2+|\tilde{e}_{\times}|^2}{2}\nonumber\\
|\tilde{h}_{\times}|^2&\rightarrow &\frac{|\tilde{e}_+|^2+|\tilde{e}_{\times}|^2}{2}.\nonumber\\
\eea

\begin{widetext}
\begin{figure}
 \begin{center}
\begin{minipage}{5cm}
\includegraphics[width=5cm,height=5cm]{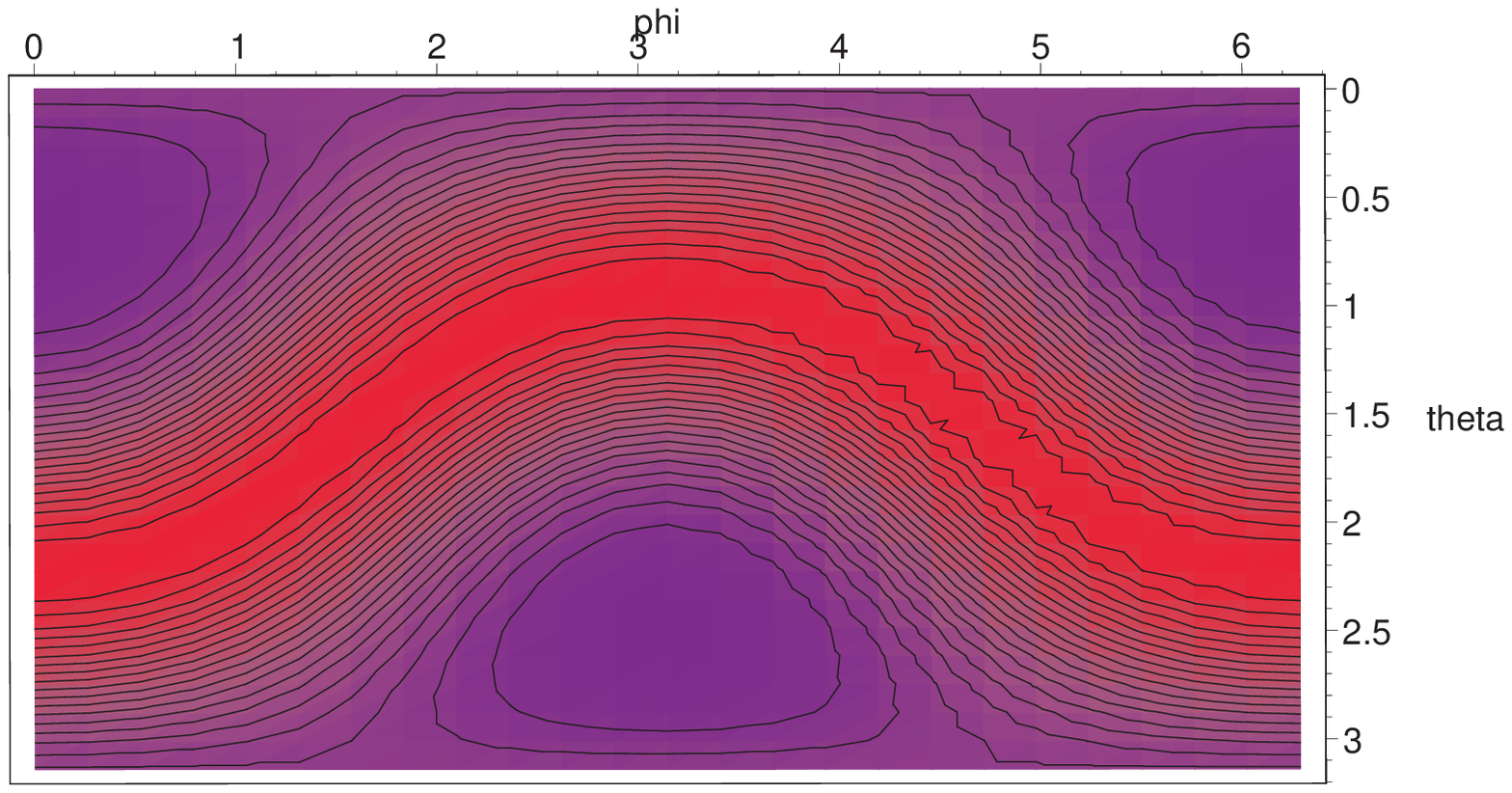}

\begin{minipage}{5cm}
\begin{center}
$A_1$
\end{center}
\end{minipage}

\end{minipage}~~\begin{minipage}{5cm}
\includegraphics[width=5cm,height=5cm]{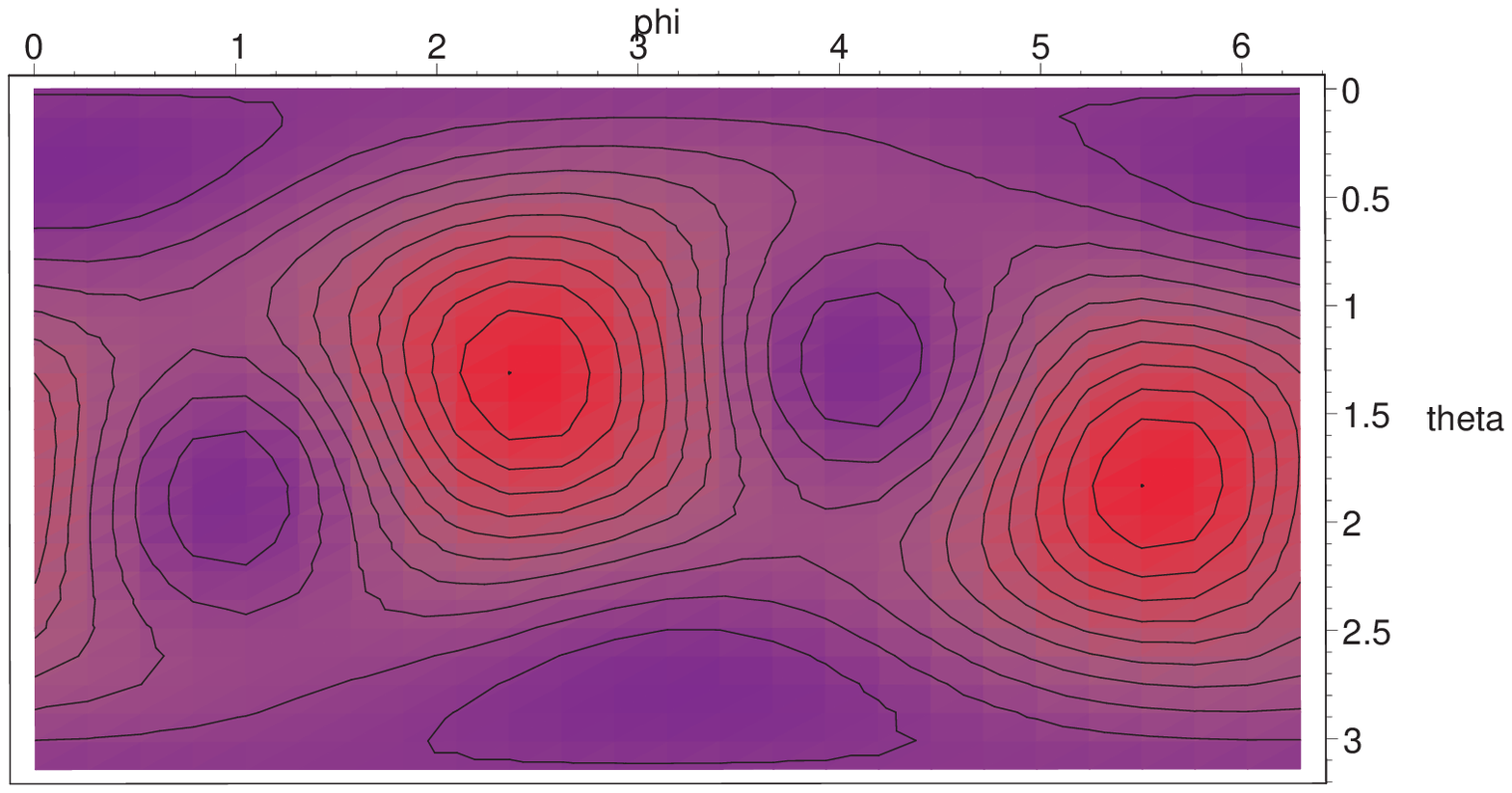}

\begin{minipage}{5cm}
\begin{center}
$A_2$
\end{center}
\end{minipage}

\end{minipage}
\end{center}

\begin{center}
\begin{minipage}{5cm}
\includegraphics[width=5cm,height=5cm]{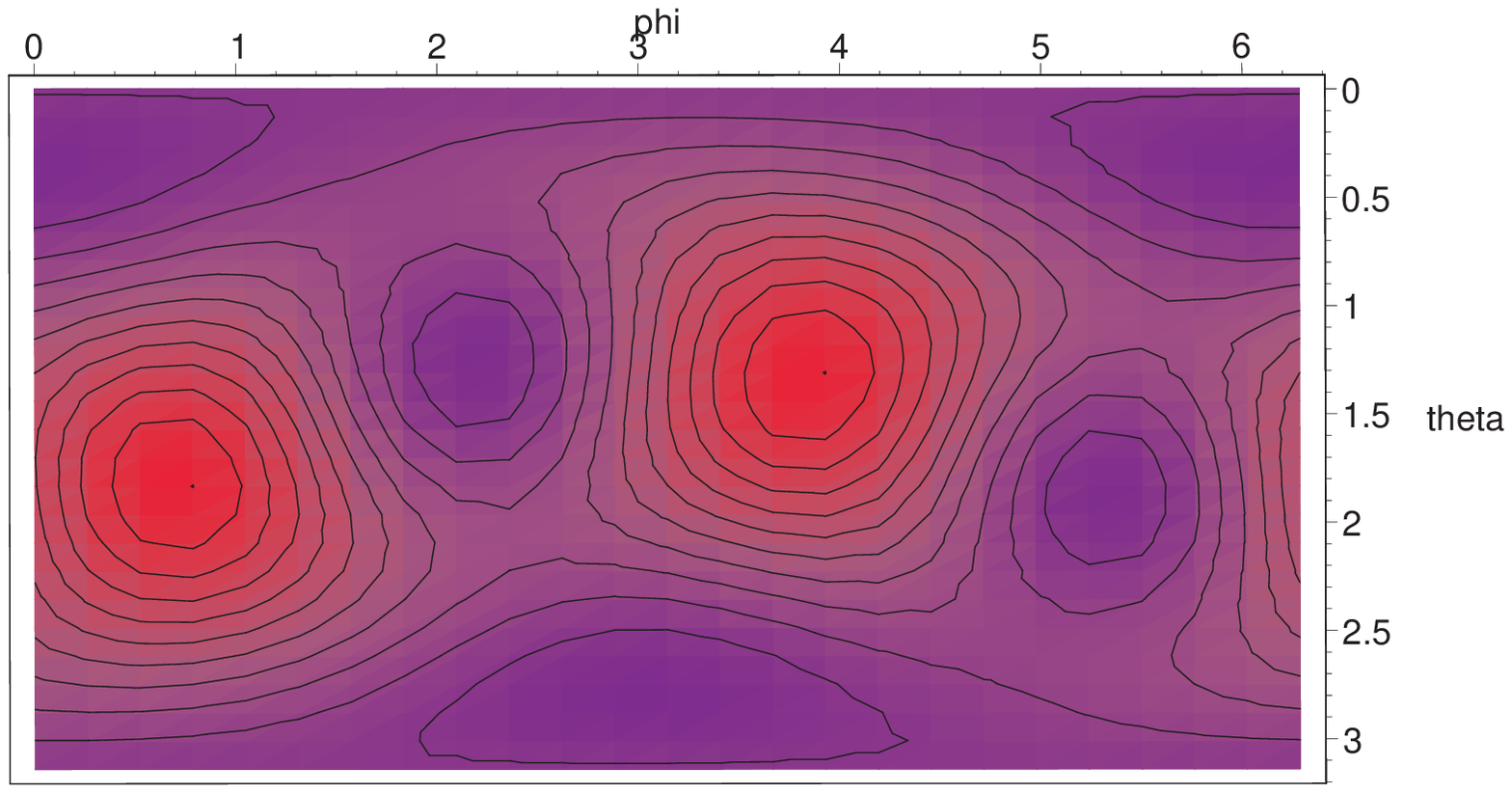}

\begin{minipage}{5cm}
\begin{center}
$A_3$
\end{center}
\end{minipage}

\end{minipage}~~\begin{minipage}{5cm}
\includegraphics[width=5cm,height=5cm]{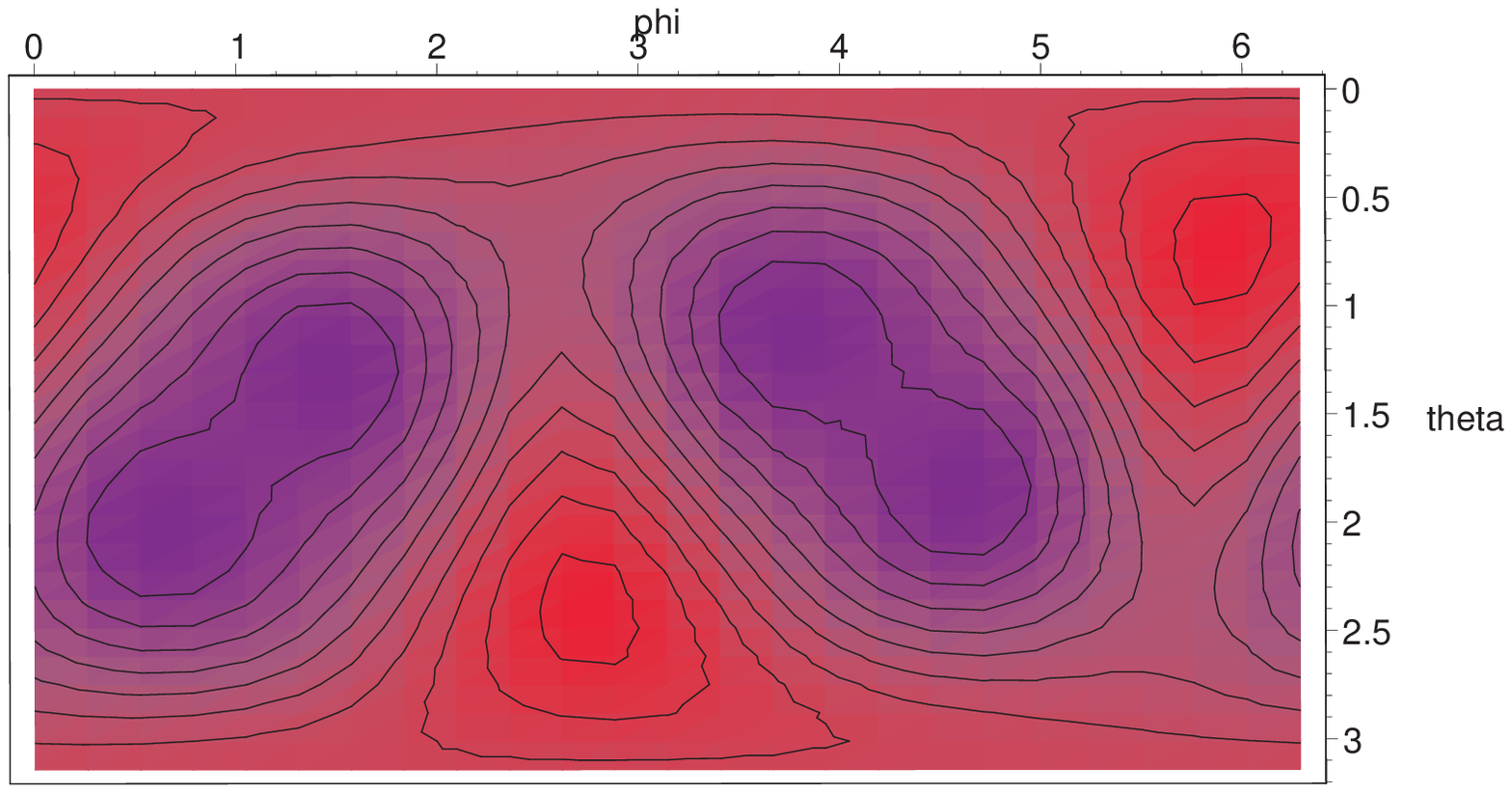}

\begin{minipage}{5cm}
\begin{center}
$A_4$
\end{center}
\end{minipage}

\end{minipage}
\end{center}

 \begin{center}
\begin{minipage}{5cm}
\includegraphics[width=5cm,height=5cm]{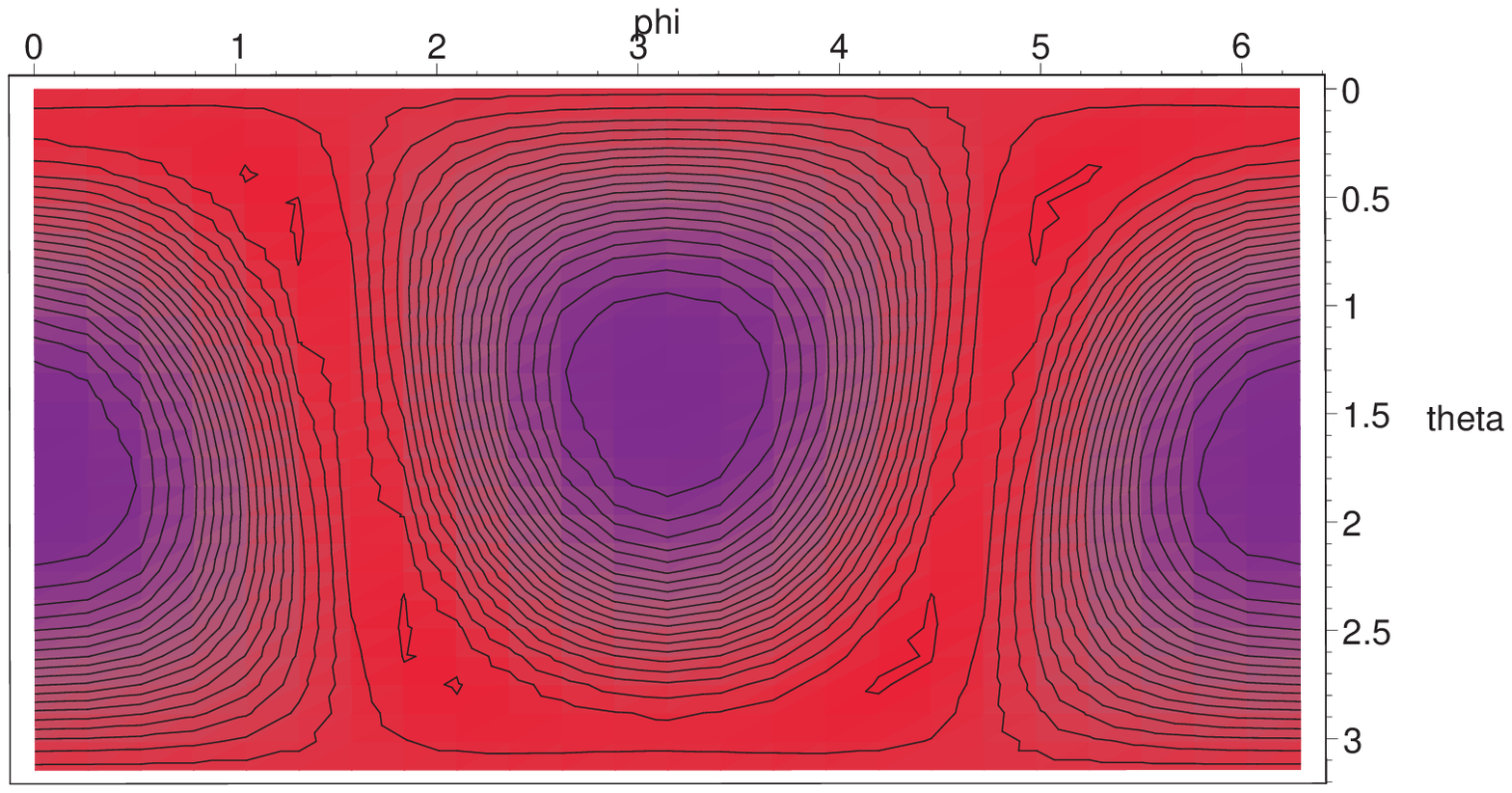}

\begin{minipage}{5cm}
\begin{center}
$A_5$
\end{center}
\end{minipage}

\end{minipage}~~\begin{minipage}{5cm}
\includegraphics[width=5cm,height=5cm]{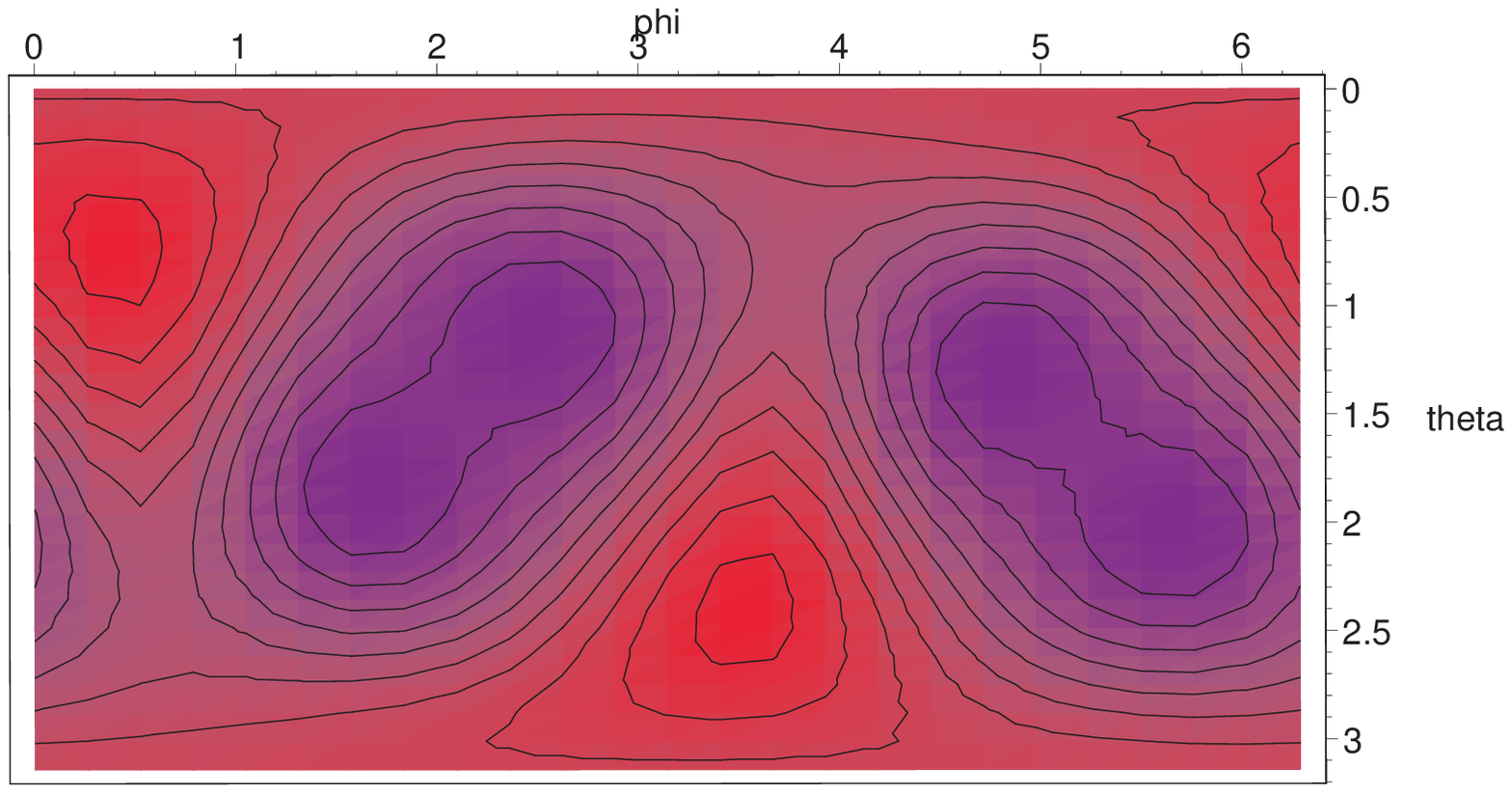}

\begin{minipage}{5cm}
\begin{center}
$A_6$
\end{center}
\end{minipage}

\end{minipage}
\end{center}
\caption{Sensitivity of the six resonators in TIGA configuration as a function of the GW arrival direction $(\theta,\phi).$ As in Figure 1, the ratio $A_{i}$ is zero where the graph is blue, and maximal where the graph is red (Color online).}
\label{fig5}
\end{figure}
\end{widetext}

The ratios $$A_i(\theta,\Phi)=\frac{|\tilde{q}_i(\omega)|^2}{\beta(\omega)(|\tilde{e}_+|^2+|\tilde{e}_{\times}|^2)/2}$$
with $\beta(\omega)=\frac{A^2a^2\omega^8}{3/5(\omega^2-\omega_+^2)^2(\omega^2-\omega_-^2)^2}$, give the sensitivities of the six resonators as a function of the direction of GW provenance.\\
The explicit expression for one of the $A_i$ previously defined is:
\begin{widetext}
\bea
&&A_2=\left[\frac{\sqrt{3}}{6}\sin^2\theta+\frac{\sqrt{3}}{18}(\sqrt{5}+1)(\cos\theta\sin\theta\cos\phi)+\frac{1}{6}(\sqrt{5}+1)(\cos\theta\sin\theta\sin\phi)\right.\nonumber\\
&&\left.-\frac{\sqrt{3}}{18}(\sqrt{5}-1)(1+\cos^2\theta)(\cos^2\phi-1/2)-\frac{1}{6}(\sqrt{5}-1)(1+\cos^2\theta)\cos\phi\sin\phi\right]^2\nonumber\\
&&+\left[-\frac{\sqrt{3}}{18}(\sqrt{5}-1)(\sin\theta\sin\phi)+\frac{1}{6}(\sqrt{5}+1)(\sin\theta\cos\phi)+\frac{\sqrt{3}}{9}(\sqrt{5}-1)\cos\theta\cos\phi\sin\phi\right.\nonumber\\
&&\left.-\frac{1}{6}(\sqrt{5}-1)\cos\theta(\cos^2\phi-\sin^2\phi)\right]^2.
\eea
\end{widetext}
The other five $A_i$ have analogous expressions. Their angular dependence is shown in Figure \ref{fig5}, while Figure 6 shows the angular dependence of their sum.

Note that the sum of the six sensitivities is no longer a constant. This is because, as we noted before, putting resonators on the surface of the sphere means destroying the original spherical symmetry. Nevertheless, the total sensitivity is never zero, so for the direction of a GW from any provenance, there is at least one resonator which will be excited significantly. Omnidirectional sensitivity is then preserved in the TIGA configuration, even if omnidirectional symmetry is broken.
 
In Figure \ref{fig7} we take into account the earth's rotation, as explained in the appendix of \cite{Flox}, for a fixed source near the galactic center. Angular sensitivities for the six resonators and for their sum are shown as a function of sidereal hour and galactic longitude. As before, we take as an example the sphere near Leiden, Holland ( latitude $l=52.16^{\circ}N$ , longitude $L=4.45^{\circ}E$).

\section{Summary and Conclusions}

In this work, the general problem of the interaction of a spherical detector with GWs has been analyzed, in both the cases of a simple elastic sphere and of a system sphere + transducers. The five responses of the quadrupole fundamental modes and of resonators motions for the TIGA configuration have been found, explicitly identifying angular dependence and, for GW radiation coming from a source in the galactic plane, sidereal time dependence as well.

We find the expected omnidirectional constant sensitivity for the ideal case of a simple sphere. 
This omnidirectional sensitivity is shown to be preserved in the case of the sphere equipped with TIGA transducers, nevertheless it is interesting  to note that it is no longer a constant. This confirms the fact that putting resonators on the surface of the sphere means choosing some favorite directions, breaking the omnidirectional initial symmetry.

For current experiments, the results of this paper allow one to predict exactly how the resonators should behave once the position of a candidate source is known. Vice versa, relations between the resonators' amplitude and $(\theta(t),\phi(t))$ can be inversed and whenever there is a statistically significative reason for thinking that a gravitational wave signal has been received, one can estimate the source direction by looking at the ratio between resonators' excitations and at the GW arrival time.   

Nevertheless, all this work is based on an ideal sphere and ideal radial moving resonators. Deviations and errors on the above predictions exist in a real spherical detector experiment. In particular, the asymmetry caused by the hole drilled through the center for the suspension system induces a degeneracy of the quadrupole five fundamental modes, which in \cite{M&J1} is estimated to be less than 1$\%$. We have not derived the consequences of this degeneracy in this paper, but it would be interesting to do that in order to better match theoretical predictions with experiments.

\begin{figure}\label{fig6}
\begin{center}

\includegraphics[width=4.5cm,height=4.5cm]{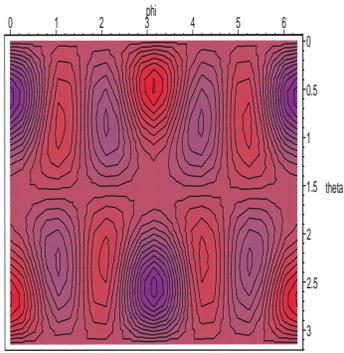}
\end{center}
\caption{The same as Figure 5 for the sum of the $A_i$ (Color online).}
\end{figure}

\begin{widetext}
\begin{figure}  
 \begin{center}
\begin{minipage}{5cm}
\includegraphics[width=5cm,height=5cm]{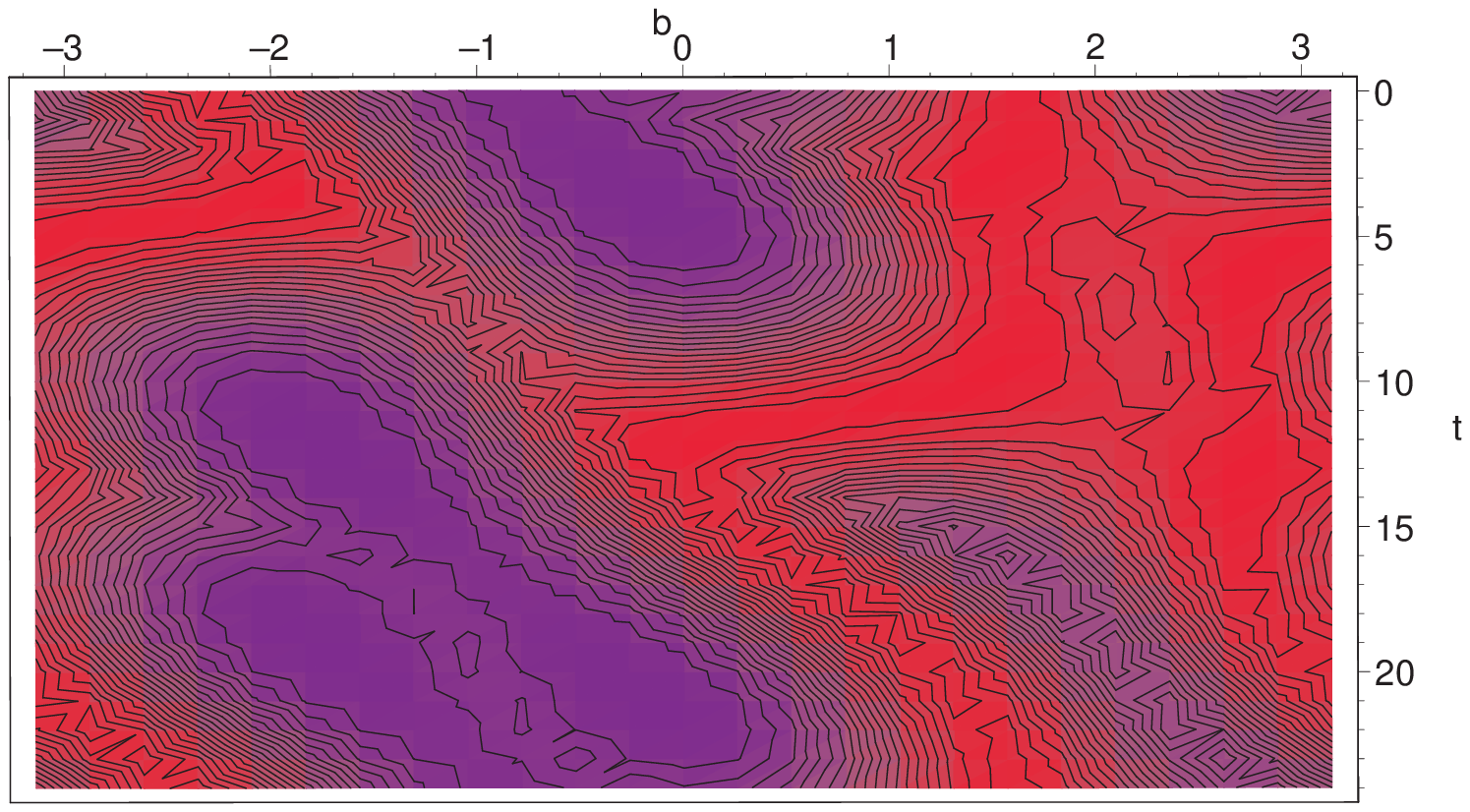}

\begin{minipage}{5cm}
\begin{center}
$A_1$
\end{center}
\end{minipage}

\end{minipage}~~\begin{minipage}{5cm}
\includegraphics[width=5cm,height=5cm]{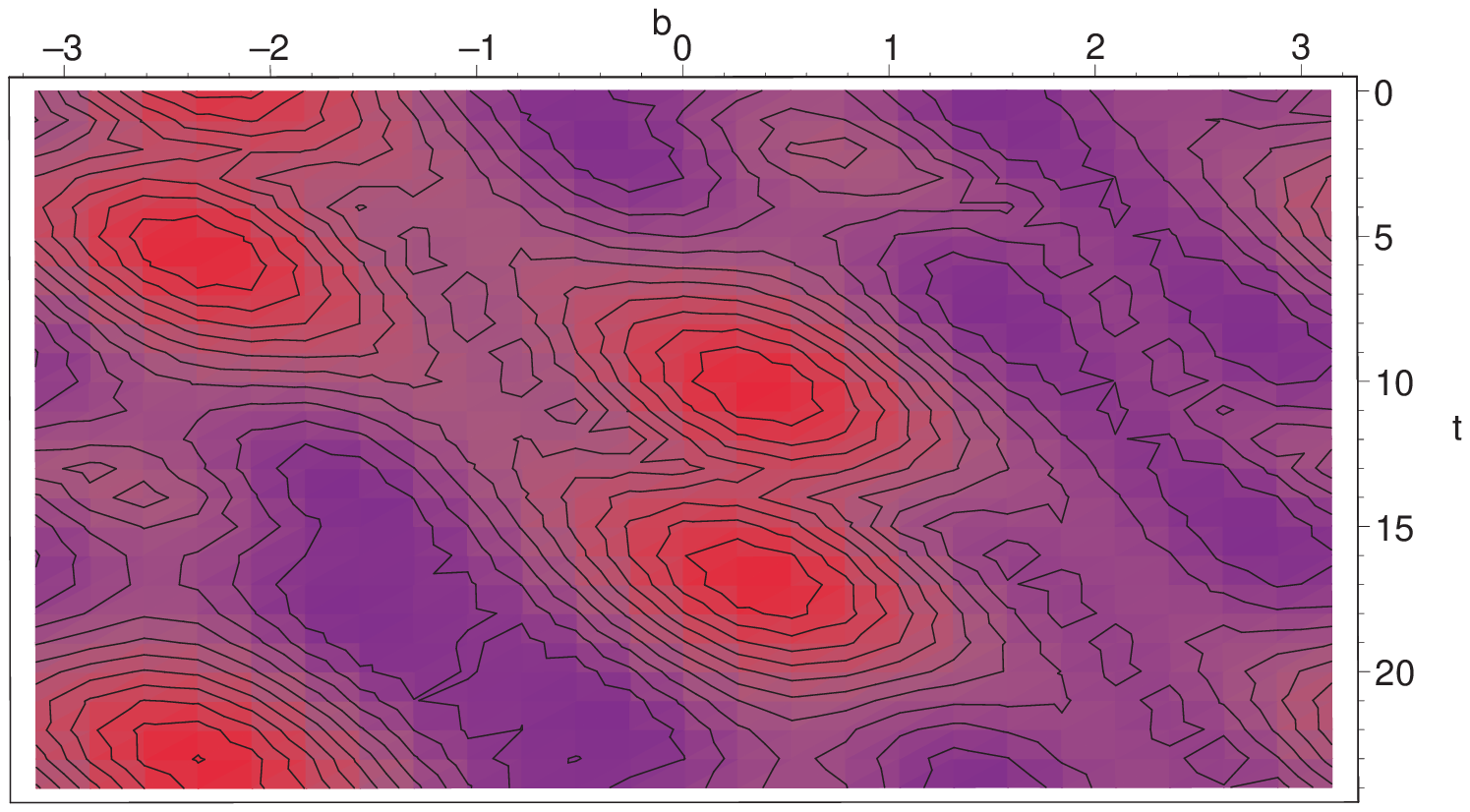}

\begin{minipage}{5cm}
\begin{center}
$A_2$
\end{center}
\end{minipage}

\end{minipage}
\end{center}

\begin{center}
\begin{minipage}{5cm}
\includegraphics[width=5cm,height=5cm]{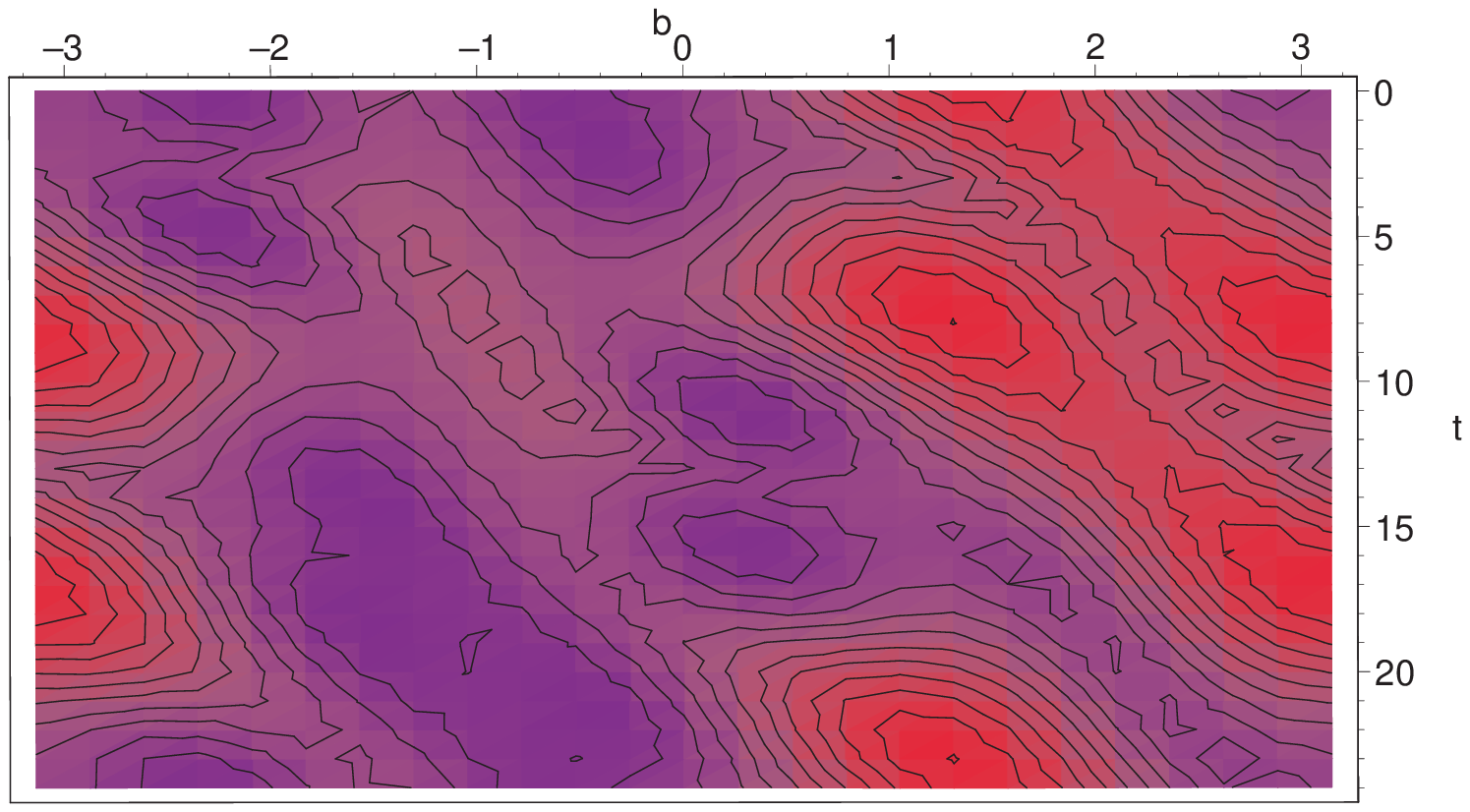}

\begin{minipage}{5cm}
\begin{center}
$A_3$
\end{center}
\end{minipage}

\end{minipage}~~\begin{minipage}{5cm}
\includegraphics[width=5cm,height=5cm]{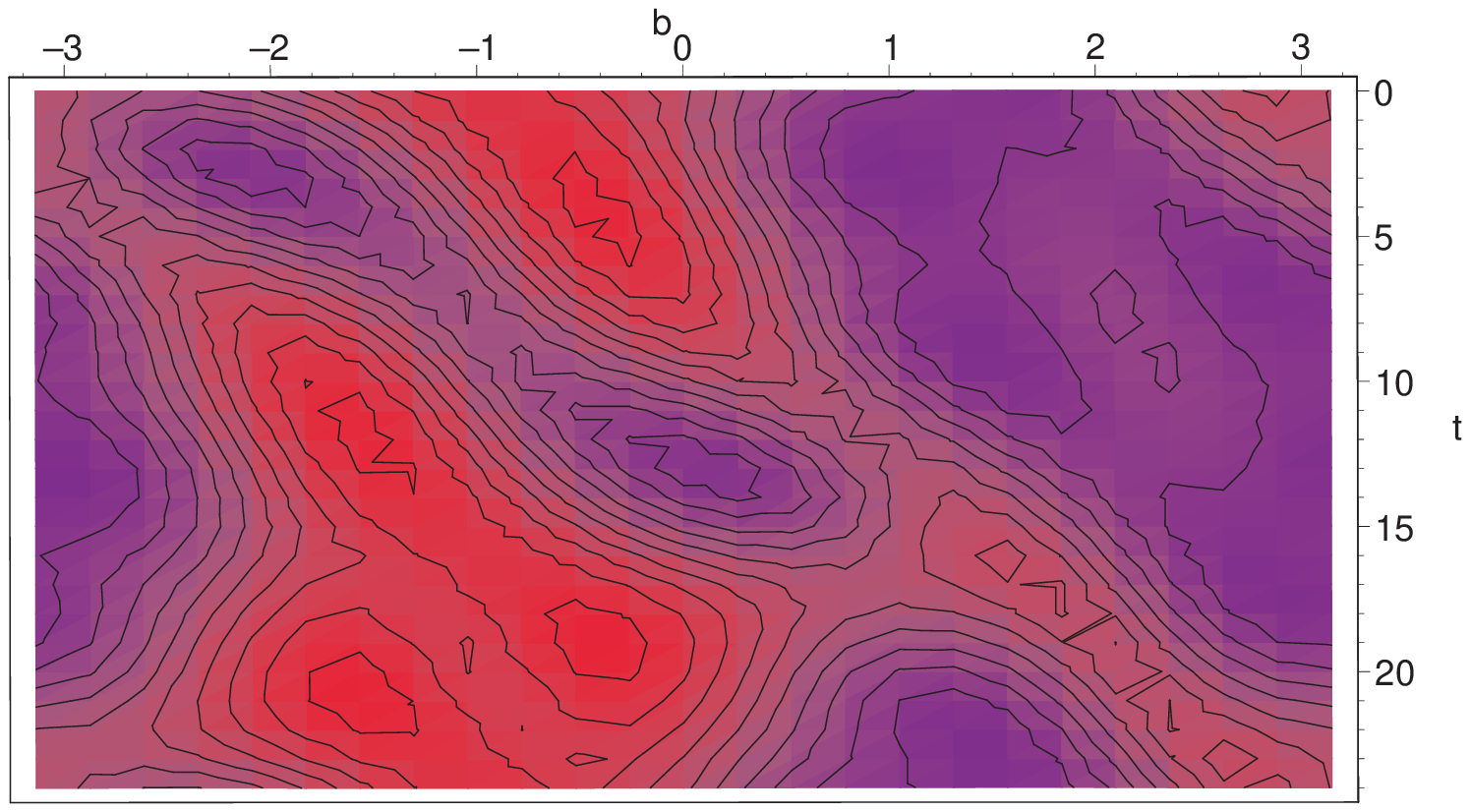}

\begin{minipage}{5cm}
\begin{center}
$A_4$
\end{center}
\end{minipage}

\end{minipage}
\end{center}

\begin{center}
\begin{minipage}{5cm}
\includegraphics[width=5cm,height=5cm]{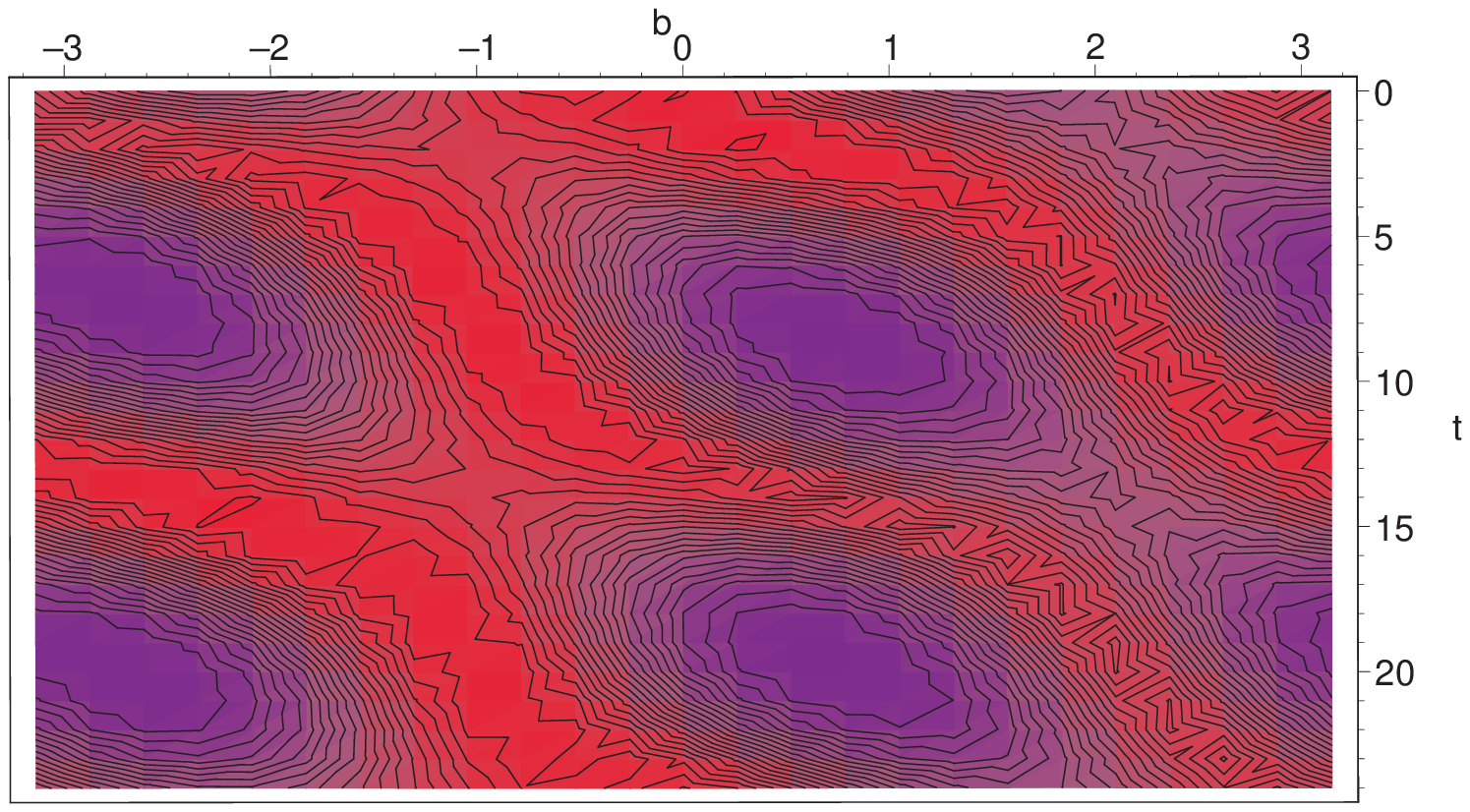}

\begin{minipage}{5cm}
\begin{center}
$A_5$
\end{center}
\end{minipage}

\end{minipage}~~\begin{minipage}{5cm}
\includegraphics[width=5cm,height=5cm]{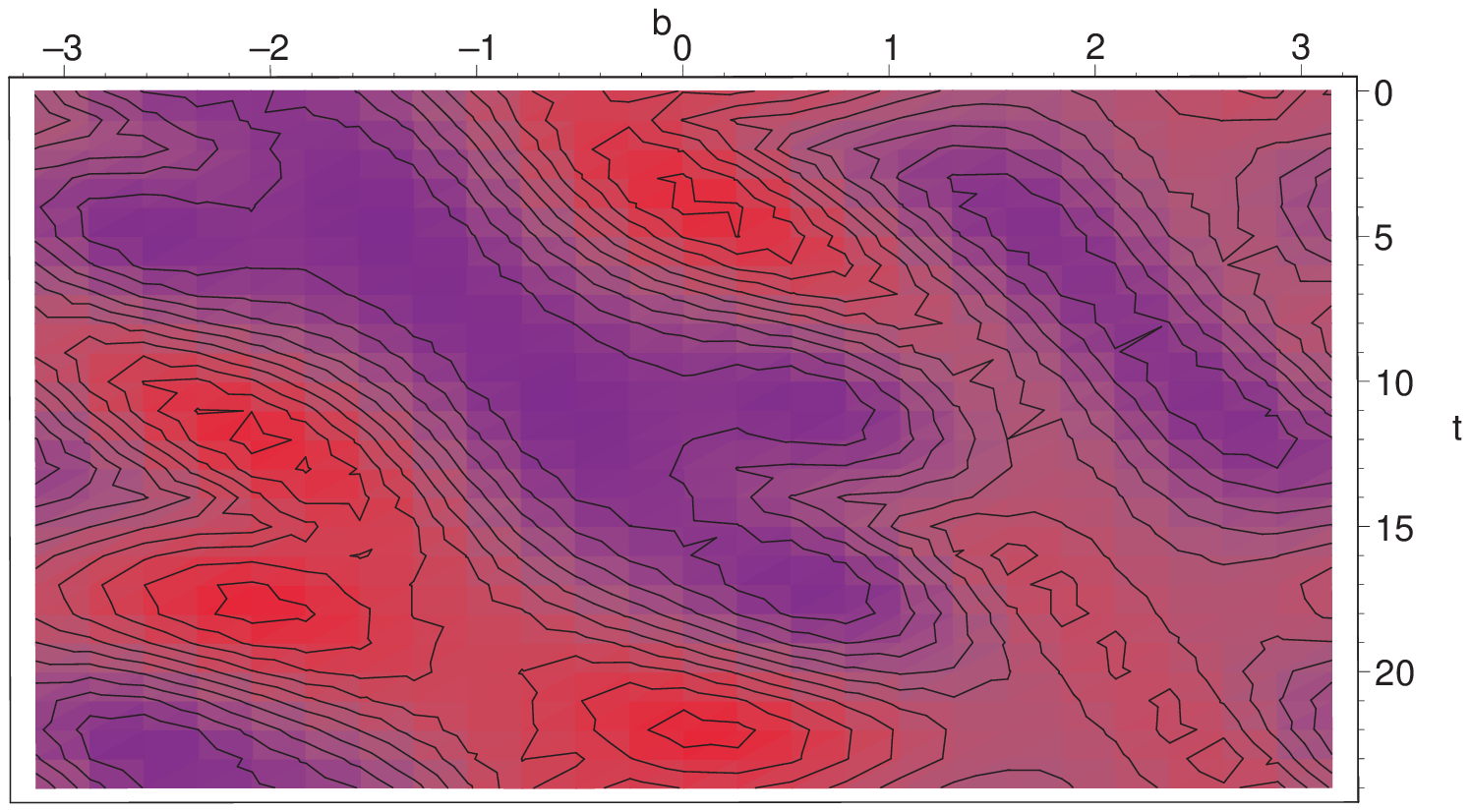}

\begin{minipage}{5cm}
\begin{center}
$A_6$
\end{center}
\end{minipage}

\end{minipage}
\end{center}
\caption{The same as Figure 5, but as a function of sidereal hours (t) and galactic longitude (b) of the GW source, for a spherical detector near Leiden, NL, $l=52.16^{\circ}N$, $L=4.45^{\circ}E$ (Color online). }
\label{fig7}
\end{figure}
\end{widetext}

\begin{figure}\label{fig8} 
\begin{center}
\begin{minipage}{5cm}
\includegraphics[width=5cm,height=5cm]{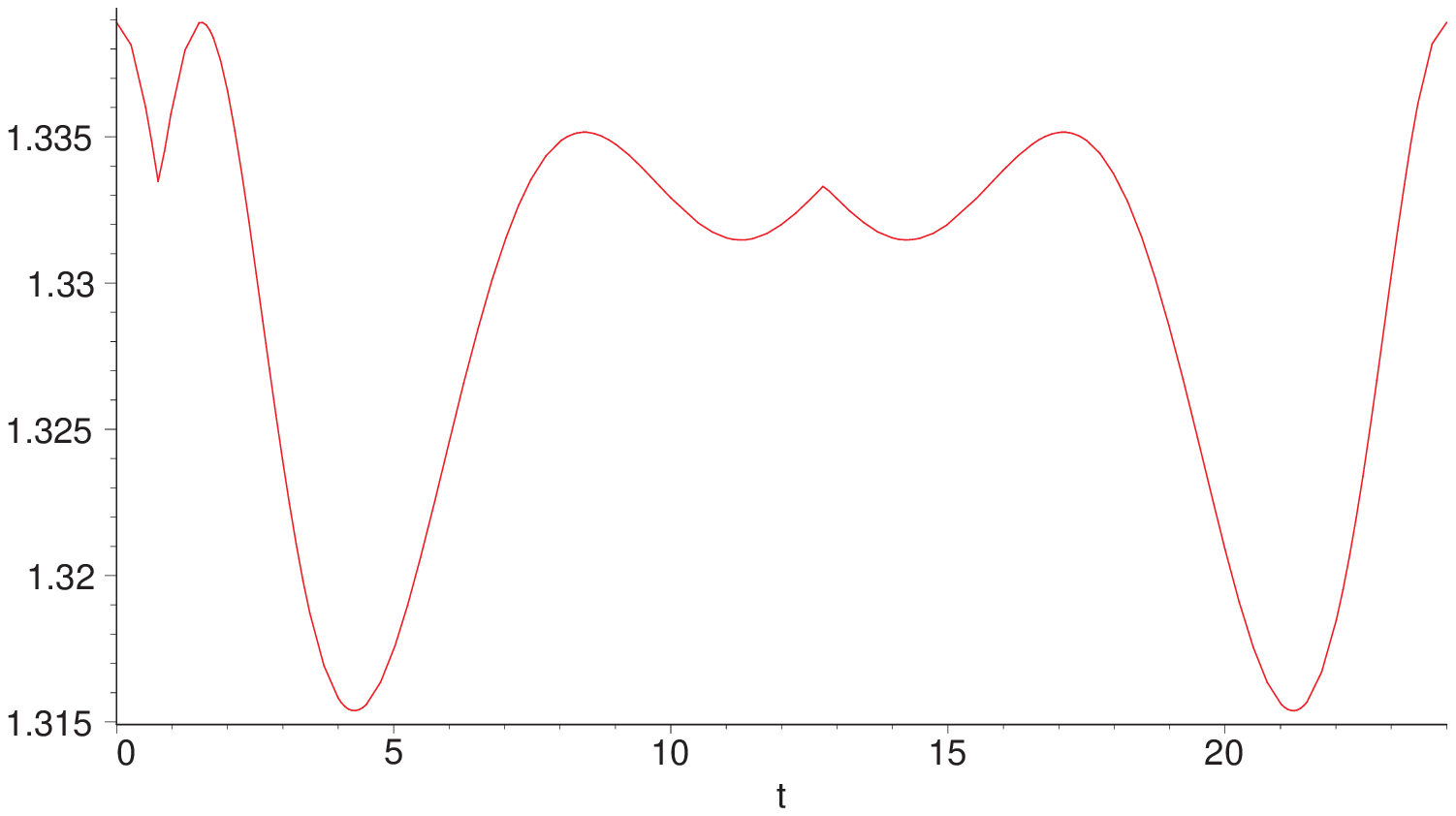}
\end{minipage}
\end{center}
\caption{Sum of the $A_i$ of Figure \ref{fig7} as a function of sidereal time for b=0, i.e. a source near the galactic center (Color online).}
\end{figure}


\begin{acknowledgments}
I would like to thank Florian Dubath and Stefano Foffa for many helpful discussions.\\
A special thank to Angela Haden, Danielle Chevalier, Andreas Malaspinas, Marti Ruiz-Altaba and Fabio Dubath.\\
This work is partially supported by the Swiss National Fund (FNS).
\end{acknowledgments}

\end{document}